\documentclass[11pt,english]{article}
\pdfoutput=1
\usepackage[T1]{fontenc}
\usepackage{amssymb}
\usepackage{amsmath}
\DeclareMathOperator{\ord}{ord}
\usepackage[unicode=true,
bookmarks=true,bookmarksnumbered=false,bookmarksopen=false,
 breaklinks=false,pdfborder={0 0 1},backref=false,colorlinks=true]
 {hyperref}
\usepackage[hang,flushmargin]{footmisc}
\usepackage{relsize}
\usepackage{color}

\usepackage{amsfonts}
\usepackage{graphicx}
\usepackage{longtable}

\usepackage{lineno}
\usepackage{tcolorbox}
\tcbuselibrary{breakable}
\usepackage{placeins}
\usepackage{float}

\DeclareFontShape{OT1}{cmr}{mx}{n}%
{<->cmr10}{}

\usepackage{tocloft}
\usepackage{etoolbox}

\hypersetup{
    colorlinks=true,
    linkcolor=blue,  
    urlcolor=blue,
    citecolor=blue
}

\makeatletter
\pretocmd{\l@section}{\hypersetup{linkcolor=black}}{}{}
\apptocmd{\l@section}{\hypersetup{linkcolor=blue}}{}{}
\pretocmd{\l@subsection}{\hypersetup{linkcolor=black}}{}{}
\apptocmd{\l@subsection}{\hypersetup{linkcolor=blue}}{}{}
\pretocmd{\l@subsubsection}{\hypersetup{linkcolor=black}}{}{}
\apptocmd{\l@subsubsection}{\hypersetup{linkcolor=blue}}{}{}
\makeatother

\usepackage{amsmath,amssymb,amsthm,mathtools}
\numberwithin{equation}{section}
\usepackage{tikz-cd}

\usepackage[normalem]{ulem}
\definecolor{JBDComment}{rgb}{0.50,0.20,0.65}

\definecolor{CCComment}{rgb}{0.00,0.50,0.30}

\definecolor{JPMComment}{rgb}{0.85,0.33,0.10}

\definecolor{RTComment}{rgb}{0.86,0.10,0.10}

\newcommand{\AuthorEdits}[1]{%
  \expandafter\newcommand\csname#1replace\endcsname[2]{\textcolor{#1Comment}{\ifstrempty{##1}{##2}{\sout{##1}\ifstrempty{##2}{}{ ##2}}}}%
}
\forcsvlist{\AuthorEdits}{JBD,CC,JPM,RT}

\newcommand{\Hom}[0]{\text{Hom}}
\newcommand{\C}[0]{\mathcal C}
\newcommand{\id}[0]{\textup{id}}

\newcommand{\zee}[1]{\mathbb Z_{#1}}

\newcommand{\Uone}{\text{U}(1)}

\newcommand{\bra}[1]{\left\langle #1 \right\rvert}
\newcommand{\ket}[1]{\left\lvert #1 \right\rangle}

\usepackage[
  backend=bibtex,
  style=phys,
  biblabel=brackets,
  sorting=none,
  eprint=true,
  doi=true,
  url=false,
  isbn=false,
  giveninits=true,
  maxbibnames=10,
]{biblatex}
\usepackage{subcaption}
\usepackage{cleveref}
\crefname{figure}{Fig.}{Figs.}
\Crefname{figure}{Fig.}{Figs.}
\crefname{equation}{Eq.}{Eqs.}
\Crefname{equation}{Eq.}{Eqs.}
\crefname{section}{Section}{Sections}
\Crefname{section}{Section}{Sections}
\crefname{subsection}{Section}{Sections}
\Crefname{subsection}{Section}{Sections}
\crefname{appendix}{Appendix}{Appendices}
\Crefname{appendix}{Appendix}{Appendices}
\AddToHook{cmd/appendix/after}{%
  \crefalias{subsection}{appendix}%
  \crefalias{subsubsection}{appendix}%
}
\AtBeginDocument{}

\tikzset{
  appendix strand/.style={
    draw=black,
    line width=0.75pt,
    line cap=round,
    line join=round
  },
  appendix crossing/.style={
    appendix strand,
    preaction={draw=white, line width=3.2pt}
  },
  appendix vertex/.style={circle, fill=black, inner sep=1.25pt},
  appendix label/.style={font=\scriptsize, inner sep=1pt},
  appendix channel/.style={font=\scriptsize, inner sep=0pt},
  appendix basis/.style={font=\scriptsize, inner sep=0pt}
}

\tikzset{
  appendix strand/.style={
    draw=black,
    line width=0.75pt,
    line cap=round,
    line join=round
  },
  appendix crossing/.style={
    appendix strand,
    preaction={draw=white, line width=3.2pt}
  },
  appendix vertex/.style={circle, fill=black, inner sep=1.25pt},
  appendix label/.style={font=\scriptsize, inner sep=1pt},
  appendix channel/.style={font=\scriptsize, inner sep=0pt},
  appendix basis/.style={font=\scriptsize, inner sep=0pt}
}

\usetikzlibrary{positioning}

\newcommand{\LeftAssociatedFusionTree}{%
  \begin{tikzpicture}[baseline=-0.52ex, x=0.58cm, y=0.54cm]
    \path[use as bounding box] (-1.48,-1.10) rectangle (1.48,2.42);
    \coordinate (LaTop) at (-1.20,2.12);
    \coordinate (LbTop) at (-0.16,2.12);
    \coordinate (LcTop) at (1.18,2.12);
    \coordinate (Lupper) at (-0.64,1.08);
    \coordinate (Llower) at (0,0);
    \coordinate (Lbottom) at (0,-0.82);
    \draw[appendix strand] (LaTop)
      .. controls (-1.20,1.70) and (-0.88,1.27) .. (Lupper);
    \draw[appendix strand] (LbTop)
      .. controls (-0.16,1.70) and (-0.39,1.27) .. (Lupper);
    \draw[appendix strand] (Lupper)
      .. controls (-0.58,0.75) and (-0.25,0.22)
      .. coordinate[pos=0.62] (LeMid) (Llower);
    \draw[appendix strand] (LcTop)
      .. controls (1.18,1.28) and (0.63,0.36) .. (Llower);
    \draw[appendix strand] (Llower) -- (Lbottom);
    \node[appendix vertex] (LupperDot) at (Lupper) {};
    \node[appendix vertex] (LlowerDot) at (Llower) {};
    \node[appendix label, above=0pt of LaTop] {$a$};
    \node[appendix label, above=0pt of LbTop] {$b$};
    \node[appendix label, above=0pt of LcTop] {$c$};
    \node[appendix basis, left=3.5pt of LupperDot] {$\beta$};
    \node[appendix channel, left=4.5pt of LeMid] {$e$};
    \node[appendix basis, right=3.5pt of LlowerDot] {$\alpha$};
    \node[appendix label, below=0pt of Lbottom] {$d$};
  \end{tikzpicture}%
}

\newcommand{\RightAssociatedFusionTree}{%
  \begin{tikzpicture}[baseline=-0.52ex, x=0.58cm, y=0.54cm]
    \path[use as bounding box] (-1.48,-1.10) rectangle (1.48,2.42);
    \coordinate (RaTop) at (-1.18,2.12);
    \coordinate (RbTop) at (0.16,2.12);
    \coordinate (RcTop) at (1.20,2.12);
    \coordinate (Rupper) at (0.64,1.08);
    \coordinate (Rlower) at (0,0);
    \coordinate (Rbottom) at (0,-0.82);
    \draw[appendix strand] (RbTop)
      .. controls (0.16,1.70) and (0.39,1.27) .. (Rupper);
    \draw[appendix strand] (RcTop)
      .. controls (1.20,1.70) and (0.88,1.27) .. (Rupper);
    \draw[appendix strand] (Rupper)
      .. controls (0.58,0.75) and (0.25,0.22)
      .. coordinate[pos=0.62] (RfMid) (Rlower);
    \draw[appendix strand] (RaTop)
      .. controls (-1.18,1.28) and (-0.63,0.36) .. (Rlower);
    \draw[appendix strand] (Rlower) -- (Rbottom);
    \node[appendix vertex] (RupperDot) at (Rupper) {};
    \node[appendix vertex] (RlowerDot) at (Rlower) {};
    \node[appendix label, above=0pt of RaTop] {$a$};
    \node[appendix label, above=0pt of RbTop] {$b$};
    \node[appendix label, above=0pt of RcTop] {$c$};
    \node[appendix basis, right=3.5pt of RupperDot] {$\nu$};
    \node[appendix channel, right=4.5pt of RfMid] {$f$};
    \node[appendix basis, left=3.5pt of RlowerDot] {$\mu$};
    \node[appendix label, below=0pt of Rbottom] {$d$};
  \end{tikzpicture}%
}

\tikzset{
  appendix strand/.style={
    draw=black,
    line width=0.75pt,
    line cap=round,
    line join=round
  },
  appendix crossing/.style={
    appendix strand,
    preaction={draw=white, line width=3.2pt}
  },
  appendix vertex/.style={circle, fill=black, inner sep=1.25pt},
  appendix label/.style={font=\scriptsize, inner sep=1pt},
  appendix channel/.style={font=\scriptsize, inner sep=0pt},
  appendix basis/.style={font=\scriptsize, inner sep=0pt}
}

\usetikzlibrary{arrows.meta,calc,positioning}

\newcommand{\PentagonChannelAt}[4]{%
  \coordinate (#1Label) at ($(#1)!5pt!#3:(#2)$);
  \node[appendix channel] at (#1Label) {$#4$};
}

\newcommand{\PentagonExternalLabels}[5]{%
  \node[appendix label, above=0pt of PTopA] {$#1$};
  \node[appendix label, above=0pt of PTopB] {$#2$};
  \node[appendix label, above=0pt of PTopC] {$#3$};
  \node[appendix label, above=0pt of PTopD] {$#4$};
  \node[appendix label, below=0pt of PBottom] {$#5$};
}

\newcommand{\PentagonLeftComb}{%
  \begin{tikzpicture}[baseline=-0.52ex, x=0.42cm, y=0.40cm]
    \path[use as bounding box] (-1.62,-1.05) rectangle (1.62,2.50);
    \coordinate (PTopA) at (-1.38,2.18);
    \coordinate (PTopB) at (-0.54,2.18);
    \coordinate (PTopC) at (0.34,2.18);
    \coordinate (PTopD) at (1.38,2.18);
    \coordinate (PVertexOne) at (-0.96,1.34);
    \coordinate (PVertexTwo) at (-0.43,0.58);
    \coordinate (PVertexThree) at (0,-0.18);
    \coordinate (PBottom) at (0,-0.78);
    \draw[appendix strand] (PTopA)
      .. controls (-1.38,1.84) and (-1.15,1.52) .. (PVertexOne);
    \draw[appendix strand] (PTopB)
      .. controls (-0.54,1.84) and (-0.75,1.52) .. (PVertexOne);
    \draw[appendix strand] (PVertexOne)
      .. controls (-0.88,1.06) and (-0.58,0.76)
      .. coordinate[pos=0.50] (PEdgeE)
         coordinate[pos=0.60] (PEdgeEForward) (PVertexTwo);
    \draw[appendix strand] (PTopC)
      .. controls (0.34,1.53) and (-0.04,0.82) .. (PVertexTwo);
    \draw[appendix strand] (PVertexTwo)
      .. controls (-0.39,0.29) and (-0.16,0.02)
      .. coordinate[pos=0.50] (PEdgeF)
         coordinate[pos=0.60] (PEdgeFForward) (PVertexThree);
    \draw[appendix strand] (PTopD)
      .. controls (1.38,1.30) and (0.58,0.18) .. (PVertexThree);
    \draw[appendix strand] (PVertexThree) -- (PBottom);
    \node[appendix vertex] at (PVertexOne) {};
    \node[appendix vertex] at (PVertexTwo) {};
    \node[appendix vertex] at (PVertexThree) {};
    \PentagonChannelAt{PEdgeE}{PEdgeEForward}{-90}{e}
    \PentagonChannelAt{PEdgeF}{PEdgeFForward}{-90}{f}
    \PentagonExternalLabels{a}{b}{c}{d}{x}
  \end{tikzpicture}%
}

\newcommand{\PentagonPairTree}{%
  \begin{tikzpicture}[baseline=-0.52ex, x=0.42cm, y=0.40cm]
    \path[use as bounding box] (-1.62,-1.05) rectangle (1.62,2.50);
    \coordinate (PTopA) at (-1.38,2.18);
    \coordinate (PTopB) at (-0.46,2.18);
    \coordinate (PTopC) at (0.46,2.18);
    \coordinate (PTopD) at (1.38,2.18);
    \coordinate (PVertexOne) at (-0.90,1.18);
    \coordinate (PVertexTwo) at (0.90,1.18);
    \coordinate (PVertexThree) at (0,-0.18);
    \coordinate (PBottom) at (0,-0.78);
    \draw[appendix strand] (PTopA)
      .. controls (-1.38,1.76) and (-1.14,1.38) .. (PVertexOne);
    \draw[appendix strand] (PTopB)
      .. controls (-0.46,1.76) and (-0.66,1.38) .. (PVertexOne);
    \draw[appendix strand] (PTopC)
      .. controls (0.46,1.76) and (0.66,1.38) .. (PVertexTwo);
    \draw[appendix strand] (PTopD)
      .. controls (1.38,1.76) and (1.14,1.38) .. (PVertexTwo);
    \draw[appendix strand] (PVertexOne)
      .. controls (-0.77,0.72) and (-0.30,0.14)
      .. coordinate[pos=0.50] (PEdgeE)
         coordinate[pos=0.60] (PEdgeEForward) (PVertexThree);
    \draw[appendix strand] (PVertexTwo)
      .. controls (0.77,0.72) and (0.30,0.14)
      .. coordinate[pos=0.50] (PEdgeG)
         coordinate[pos=0.60] (PEdgeGForward) (PVertexThree);
    \draw[appendix strand] (PVertexThree) -- (PBottom);
    \node[appendix vertex] at (PVertexOne) {};
    \node[appendix vertex] at (PVertexTwo) {};
    \node[appendix vertex] at (PVertexThree) {};
    \PentagonChannelAt{PEdgeE}{PEdgeEForward}{-90}{e}
    \PentagonChannelAt{PEdgeG}{PEdgeGForward}{90}{g}
    \PentagonExternalLabels{a}{b}{c}{d}{x}
  \end{tikzpicture}%
}

\newcommand{\PentagonLowerLeftTree}{%
  \begin{tikzpicture}[baseline=-0.52ex, x=0.42cm, y=0.40cm]
    \path[use as bounding box] (-1.62,-1.05) rectangle (1.62,2.50);
    \coordinate (PTopA) at (-1.38,2.18);
    \coordinate (PTopB) at (-0.42,2.18);
    \coordinate (PTopC) at (0.46,2.18);
    \coordinate (PTopD) at (1.38,2.18);
    \coordinate (PVertexOne) at (0.02,1.34);
    \coordinate (PVertexTwo) at (-0.48,0.58);
    \coordinate (PVertexThree) at (0,-0.18);
    \coordinate (PBottom) at (0,-0.78);
    \draw[appendix strand] (PTopB)
      .. controls (-0.42,1.84) and (-0.20,1.52) .. (PVertexOne);
    \draw[appendix strand] (PTopC)
      .. controls (0.46,1.84) and (0.24,1.52) .. (PVertexOne);
    \draw[appendix strand] (PVertexOne)
      .. controls (-0.04,1.06) and (-0.34,0.76)
      .. coordinate[pos=0.50] (PEdgeK)
         coordinate[pos=0.60] (PEdgeKForward) (PVertexTwo);
    \draw[appendix strand] (PTopA)
      .. controls (-1.38,1.46) and (-0.86,0.78) .. (PVertexTwo);
    \draw[appendix strand] (PVertexTwo)
      .. controls (-0.43,0.29) and (-0.17,0.02)
      .. coordinate[pos=0.50] (PEdgeF)
         coordinate[pos=0.60] (PEdgeFForward) (PVertexThree);
    \draw[appendix strand] (PTopD)
      .. controls (1.38,1.30) and (0.58,0.18) .. (PVertexThree);
    \draw[appendix strand] (PVertexThree) -- (PBottom);
    \node[appendix vertex] at (PVertexOne) {};
    \node[appendix vertex] at (PVertexTwo) {};
    \node[appendix vertex] at (PVertexThree) {};
    \PentagonChannelAt{PEdgeK}{PEdgeKForward}{90}{k}
    \PentagonChannelAt{PEdgeF}{PEdgeFForward}{-90}{f}
    \PentagonExternalLabels{a}{b}{c}{d}{x}
  \end{tikzpicture}%
}

\newcommand{\PentagonLowerRightTree}{%
  \begin{tikzpicture}[baseline=-0.52ex, x=0.42cm, y=0.40cm]
    \path[use as bounding box] (-1.62,-1.05) rectangle (1.62,2.50);
    \coordinate (PTopA) at (-1.38,2.18);
    \coordinate (PTopB) at (-0.46,2.18);
    \coordinate (PTopC) at (0.42,2.18);
    \coordinate (PTopD) at (1.38,2.18);
    \coordinate (PVertexOne) at (-0.02,1.34);
    \coordinate (PVertexTwo) at (0.48,0.58);
    \coordinate (PVertexThree) at (0,-0.18);
    \coordinate (PBottom) at (0,-0.78);
    \draw[appendix strand] (PTopB)
      .. controls (-0.46,1.84) and (-0.24,1.52) .. (PVertexOne);
    \draw[appendix strand] (PTopC)
      .. controls (0.42,1.84) and (0.20,1.52) .. (PVertexOne);
    \draw[appendix strand] (PVertexOne)
      .. controls (0.04,1.06) and (0.34,0.76)
      .. coordinate[pos=0.50] (PEdgeK)
         coordinate[pos=0.60] (PEdgeKForward) (PVertexTwo);
    \draw[appendix strand] (PTopD)
      .. controls (1.38,1.46) and (0.86,0.78) .. (PVertexTwo);
    \draw[appendix strand] (PVertexTwo)
      .. controls (0.43,0.29) and (0.17,0.02)
      .. coordinate[pos=0.50] (PEdgeH)
         coordinate[pos=0.60] (PEdgeHForward) (PVertexThree);
    \draw[appendix strand] (PTopA)
      .. controls (-1.38,1.30) and (-0.58,0.18) .. (PVertexThree);
    \draw[appendix strand] (PVertexThree) -- (PBottom);
    \node[appendix vertex] at (PVertexOne) {};
    \node[appendix vertex] at (PVertexTwo) {};
    \node[appendix vertex] at (PVertexThree) {};
    \PentagonChannelAt{PEdgeK}{PEdgeKForward}{-90}{k}
    \PentagonChannelAt{PEdgeH}{PEdgeHForward}{90}{h}
    \PentagonExternalLabels{a}{b}{c}{d}{x}
  \end{tikzpicture}%
}

\newcommand{\PentagonRightComb}{%
  \begin{tikzpicture}[baseline=-0.52ex, x=0.42cm, y=0.40cm]
    \path[use as bounding box] (-1.62,-1.05) rectangle (1.62,2.50);
    \coordinate (PTopA) at (-1.38,2.18);
    \coordinate (PTopB) at (-0.34,2.18);
    \coordinate (PTopC) at (0.54,2.18);
    \coordinate (PTopD) at (1.38,2.18);
    \coordinate (PVertexOne) at (0.96,1.34);
    \coordinate (PVertexTwo) at (0.43,0.58);
    \coordinate (PVertexThree) at (0,-0.18);
    \coordinate (PBottom) at (0,-0.78);
    \draw[appendix strand] (PTopC)
      .. controls (0.54,1.84) and (0.75,1.52) .. (PVertexOne);
    \draw[appendix strand] (PTopD)
      .. controls (1.38,1.84) and (1.15,1.52) .. (PVertexOne);
    \draw[appendix strand] (PVertexOne)
      .. controls (0.88,1.06) and (0.58,0.76)
      .. coordinate[pos=0.50] (PEdgeG)
         coordinate[pos=0.60] (PEdgeGForward) (PVertexTwo);
    \draw[appendix strand] (PTopB)
      .. controls (-0.34,1.53) and (0.04,0.82) .. (PVertexTwo);
    \draw[appendix strand] (PVertexTwo)
      .. controls (0.39,0.29) and (0.16,0.02)
      .. coordinate[pos=0.50] (PEdgeH)
         coordinate[pos=0.60] (PEdgeHForward) (PVertexThree);
    \draw[appendix strand] (PTopA)
      .. controls (-1.38,1.30) and (-0.58,0.18) .. (PVertexThree);
    \draw[appendix strand] (PVertexThree) -- (PBottom);
    \node[appendix vertex] at (PVertexOne) {};
    \node[appendix vertex] at (PVertexTwo) {};
    \node[appendix vertex] at (PVertexThree) {};
    \PentagonChannelAt{PEdgeG}{PEdgeGForward}{90}{g}
    \PentagonChannelAt{PEdgeH}{PEdgeHForward}{90}{h}
    \PentagonExternalLabels{a}{b}{c}{d}{x}
  \end{tikzpicture}%
}

\newcommand{\PentagonEdge}[6]{%
  \coordinate (Pentagon#1#2Start) at
    ($(Pentagon#1.center)!1.05cm!(Pentagon#2.center)$);
  \coordinate (Pentagon#1#2End) at
    ($(Pentagon#2.center)!1.05cm!(Pentagon#1.center)$);
  \coordinate (Pentagon#1#2Mid) at
    ($(Pentagon#1.center)!0.50!(Pentagon#2.center)$);
  \coordinate (Pentagon#1#2LabelAnchor) at
    ($(Pentagon#1#2Mid)+(#3:0.18cm)$);
  \draw[appendix map] (Pentagon#1#2Start) -- (Pentagon#1#2End);
  \node[appendix map label, anchor=#4, yshift=#5]
    at (Pentagon#1#2LabelAnchor) {#6};
}

\newcommand{\PentagonCoherenceDiagram}{%
  \begin{tikzpicture}[
    appendix map/.style={
      draw=black,
      line width=0.70pt,
      -{Stealth[length=4.2pt,width=3.3pt]}
    },
    appendix map label/.style={
      font=\small,
      inner sep=0pt
    }
  ]
    \node[inner sep=0pt] (PentagonA) at (162:2.55cm)
      {\PentagonLeftComb};
    \node[inner sep=0pt] (PentagonB) at (90:2.55cm)
      {\PentagonPairTree};
    \node[inner sep=0pt] (PentagonC) at (18:2.55cm)
      {\PentagonRightComb};
    \node[inner sep=0pt] (PentagonD) at (-54:2.55cm)
      {\PentagonLowerRightTree};
    \node[inner sep=0pt] (PentagonE) at (-126:2.55cm)
      {\PentagonLowerLeftTree};

    \PentagonEdge{A}{B}{126}{south east}{0pt}{$[F_x^{ecd}]_{fg}$}
    \PentagonEdge{B}{C}{54}{south west}{0pt}{$[F_x^{abg}]_{eh}$}
    \PentagonEdge{A}{E}{-162}{north east}{2pt}{$[F_f^{abc}]_{ek}$}
    \PentagonEdge{E}{D}{-90}{north}{0pt}{$[F_x^{akd}]_{fh}$}
    \PentagonEdge{D}{C}{-18}{north west}{2pt}{$[F_h^{bcd}]_{kg}$}
  \end{tikzpicture}%
}

\tikzset{
  appendix strand/.style={
    draw=black,
    line width=0.75pt,
    line cap=round,
    line join=round
  },
  appendix crossing/.style={
    appendix strand,
    preaction={draw=white, line width=3.2pt}
  },
  appendix vertex/.style={circle, fill=black, inner sep=1.25pt},
  appendix label/.style={font=\scriptsize, inner sep=1pt},
  appendix channel/.style={font=\scriptsize, inner sep=0pt},
  appendix basis/.style={font=\scriptsize, inner sep=0pt}
}

\usetikzlibrary{positioning}

\newcommand{\BraidedSplittingTree}{%
  \begin{tikzpicture}[baseline=-0.52ex, x=0.58cm, y=0.54cm]
    \path[use as bounding box] (-1.15,-1.10) rectangle (1.15,2.25);
    \coordinate (BleftTop) at (-0.60,1.98);
    \coordinate (BrightTop) at (0.60,1.98);
    \coordinate (Bvertex) at (0,-0.35);
    \coordinate (BleftStart) at (-0.60,0.58);
    \coordinate (BrightStart) at (0.60,0.58);
    \coordinate (Bbottom) at (0,-0.92);
    \draw[appendix strand] (Bvertex)
      .. controls (-0.16,-0.12) and (-0.60,0.30) .. (BleftStart)
      .. controls (-0.60,0.98) and (0.60,1.26) .. (0.60,1.66)
      -- (BrightTop);
    \draw[appendix strand] (Bvertex)
      .. controls (0.16,-0.12) and (0.60,0.30) .. (BrightStart)
      .. controls (0.60,0.98) and (-0.60,1.26) .. (-0.60,1.66)
      -- (BleftTop);
    \draw[appendix crossing] (-0.31000,0.93500)
      .. controls (-0.12200,1.06020) and (0.12200,1.17980)
      .. (0.31000,1.30500);
    \draw[appendix strand] (Bbottom) -- (Bvertex);
    \node[appendix vertex] (BvertexDot) at (Bvertex) {};
    \node[appendix label, above=0pt of BleftTop] {$a$};
    \node[appendix label, above=0pt of BrightTop] {$b$};
    \node[appendix basis, left=3.5pt of BvertexDot] {$\alpha$};
    \node[appendix label, below=0pt of Bbottom] {$c$};
  \end{tikzpicture}%
}

\newcommand{\ReorderedSplittingTree}{%
  \begin{tikzpicture}[baseline=-0.52ex, x=0.58cm, y=0.54cm]
    \path[use as bounding box] (-1.15,-1.10) rectangle (1.15,2.25);
    \coordinate (SleftTop) at (-0.72,1.92);
    \coordinate (SrightTop) at (0.72,1.92);
    \coordinate (Svertex) at (0,-0.35);
    \coordinate (Sbottom) at (0,-0.92);
    \draw[appendix strand] (Svertex)
      .. controls (-0.24,0.02) and (-0.72,1.10) .. (SleftTop);
    \draw[appendix strand] (Svertex)
      .. controls (0.24,0.02) and (0.72,1.10) .. (SrightTop);
    \draw[appendix strand] (Sbottom) -- (Svertex);
    \node[appendix vertex] (SvertexDot) at (Svertex) {};
    \node[appendix label, above=0pt of SleftTop] {$a$};
    \node[appendix label, above=0pt of SrightTop] {$b$};
    \node[appendix basis, right=3.5pt of SvertexDot] {$\beta$};
    \node[appendix label, below=0pt of Sbottom] {$c$};
  \end{tikzpicture}%
}

\tikzset{
  appendix strand/.style={
    draw=black,
    line width=0.75pt,
    line cap=round,
    line join=round
  },
  appendix crossing/.style={
    appendix strand,
    preaction={draw=white, line width=3.2pt}
  },
  appendix vertex/.style={circle, fill=black, inner sep=1.25pt},
  appendix label/.style={font=\scriptsize, inner sep=1pt},
  appendix channel/.style={font=\scriptsize, inner sep=0pt},
  appendix basis/.style={font=\scriptsize, inner sep=0pt}
}

\usetikzlibrary{arrows.meta,calc,positioning}

\newcommand{\HexagonChannelAt}[4]{%
  \coordinate (#1Label) at ($(#1)!5pt!#3:(#2)$);
  \node[appendix channel] at (#1Label) {$#4$};
}

\newcommand{\HexagonLeftTree}[5]{%
  \begin{tikzpicture}[baseline=-0.52ex, x=0.48cm, y=0.45cm]
    \path[use as bounding box] (-1.48,-1.10) rectangle (1.48,2.42);
    \coordinate (HLaTop) at (-1.20,2.12);
    \coordinate (HLbTop) at (-0.16,2.12);
    \coordinate (HLcTop) at (1.18,2.12);
    \coordinate (HLupper) at (-0.64,1.08);
    \coordinate (HLlower) at (0,0);
    \coordinate (HLbottom) at (0,-0.82);
    \draw[appendix strand] (HLaTop)
      .. controls (-1.20,1.70) and (-0.88,1.27) .. (HLupper);
    \draw[appendix strand] (HLbTop)
      .. controls (-0.16,1.70) and (-0.39,1.27) .. (HLupper);
    \draw[appendix strand] (HLupper)
      .. controls (-0.58,0.75) and (-0.25,0.22)
      .. coordinate[pos=0.50] (HLEdge)
         coordinate[pos=0.60] (HLEdgeForward) (HLlower);
    \draw[appendix strand] (HLcTop)
      .. controls (1.18,1.28) and (0.63,0.36) .. (HLlower);
    \draw[appendix strand] (HLlower) -- (HLbottom);
    \node[appendix vertex] at (HLupper) {};
    \node[appendix vertex] at (HLlower) {};
    \node[appendix label, above=0pt of HLaTop] {$#1$};
    \node[appendix label, above=0pt of HLbTop] {$#2$};
    \node[appendix label, above=0pt of HLcTop] {$#3$};
    \HexagonChannelAt{HLEdge}{HLEdgeForward}{-90}{#4}
    \node[appendix label, below=0pt of HLbottom] {$#5$};
  \end{tikzpicture}%
}

\newcommand{\HexagonRightTree}[5]{%
  \begin{tikzpicture}[baseline=-0.52ex, x=0.48cm, y=0.45cm]
    \path[use as bounding box] (-1.48,-1.10) rectangle (1.48,2.42);
    \coordinate (HRaTop) at (-1.18,2.12);
    \coordinate (HRbTop) at (0.16,2.12);
    \coordinate (HRcTop) at (1.20,2.12);
    \coordinate (HRupper) at (0.64,1.08);
    \coordinate (HRlower) at (0,0);
    \coordinate (HRbottom) at (0,-0.82);
    \draw[appendix strand] (HRbTop)
      .. controls (0.16,1.70) and (0.39,1.27) .. (HRupper);
    \draw[appendix strand] (HRcTop)
      .. controls (1.20,1.70) and (0.88,1.27) .. (HRupper);
    \draw[appendix strand] (HRupper)
      .. controls (0.58,0.75) and (0.25,0.22)
      .. coordinate[pos=0.50] (HREdge)
         coordinate[pos=0.60] (HREdgeForward) (HRlower);
    \draw[appendix strand] (HRaTop)
      .. controls (-1.18,1.28) and (-0.63,0.36) .. (HRlower);
    \draw[appendix strand] (HRlower) -- (HRbottom);
    \node[appendix vertex] at (HRupper) {};
    \node[appendix vertex] at (HRlower) {};
    \node[appendix label, above=0pt of HRaTop] {$#1$};
    \node[appendix label, above=0pt of HRbTop] {$#2$};
    \node[appendix label, above=0pt of HRcTop] {$#3$};
    \HexagonChannelAt{HREdge}{HREdgeForward}{90}{#4}
    \node[appendix label, below=0pt of HRbottom] {$#5$};
  \end{tikzpicture}%
}

\newcommand{\HexagonTopLeftBraidedTree}[1]{%
  \begin{tikzpicture}[baseline=-0.52ex, x=0.48cm, y=0.45cm]
    \path[use as bounding box] (-1.48,-1.10) rectangle (1.48,2.42);
    \coordinate (HTLaTop) at (-1.20,2.12);
    \coordinate (HTLbTop) at (0,2.12);
    \coordinate (HTLcTop) at (1.20,2.12);
    \coordinate (HTLupper) at (-0.64,1.08);
    \coordinate (HTLlower) at (0,0);
    \coordinate (HTLbottom) at (0,-0.82);
    \coordinate (HTLaTurn) at (-0.31,1.60);
    \coordinate (HTLbTurn) at (-0.95,1.60);
    \draw[appendix strand] (HTLlower)
      .. controls (0.24,0.36) and (0.77,1.28) .. (HTLcTop);
    \draw[appendix strand] (HTLupper)
      .. controls (-0.88,1.28) and (-1.02,1.45) .. (HTLbTurn)
      .. controls (-0.84,1.85) and (-0.22,1.90) .. (HTLbTop);
    \draw[appendix strand] (HTLupper)
      .. controls (-0.40,1.28) and (-0.24,1.45) .. (HTLaTurn)
      .. controls (-0.45,1.85) and (-1.00,1.90) .. (HTLaTop);
    \ifnum#1=1
      \draw[appendix crossing] (-0.33677,1.63959)
        .. controls (-0.45641,1.78943) and (-0.73356,1.86243)
        .. (-0.95549,1.96213);
    \else
      \draw[appendix crossing] (-0.92354,1.64473)
        .. controls (-0.81724,1.78601) and (-0.53452,1.85843)
        .. (-0.29656,1.95191);
    \fi
    \draw[appendix strand] (HTLupper)
      .. controls (-0.58,0.75) and (-0.25,0.22)
      .. coordinate[pos=0.50] (HTLEdge)
         coordinate[pos=0.60] (HTLEdgeForward) (HTLlower);
    \draw[appendix strand] (HTLbottom) -- (HTLlower);
    \node[appendix vertex] at (HTLupper) {};
    \node[appendix vertex] at (HTLlower) {};
    \node[appendix label, above=0pt of HTLaTop] {$a$};
    \node[appendix label, above=0pt of HTLbTop] {$b$};
    \node[appendix label, above=0pt of HTLcTop] {$c$};
    \HexagonChannelAt{HTLEdge}{HTLEdgeForward}{-90}{e}
    \node[appendix label, below=0pt of HTLbottom] {$d$};
  \end{tikzpicture}%
}

\newcommand{\HexagonTopRightBraidedTree}[1]{%
  \begin{tikzpicture}[baseline=-0.52ex, x=0.48cm, y=0.45cm]
    \path[use as bounding box] (-1.48,-1.10) rectangle (1.48,2.42);
    \coordinate (HTRaTop) at (-1.20,2.12);
    \coordinate (HTRbTop) at (0,2.12);
    \coordinate (HTRcTop) at (1.20,2.12);
    \coordinate (HTRupper) at (0.64,1.08);
    \coordinate (HTRlower) at (0,0);
    \coordinate (HTRbottom) at (0,-0.82);
    \coordinate (HTRaTurn) at (-0.05,1.52);
    \coordinate (HTRbTurn) at (-0.78,1.42);
    \draw[appendix strand] (HTRupper)
      .. controls (0.86,1.34) and (1.16,1.74) .. (HTRcTop);
    \draw[appendix strand] (HTRlower)
      .. controls (-0.45,0.45) and (-0.85,1.00) .. (HTRbTurn)
      .. controls (-0.72,1.72) and (-0.22,1.88) .. (HTRbTop);
    \draw[appendix strand] (HTRupper)
      .. controls (0.42,1.28) and (0.08,1.38) .. (HTRaTurn)
      .. controls (-0.30,1.78) and (-0.90,1.88) .. (HTRaTop);
    \ifnum#1=1
      \draw[appendix crossing] (-0.20510,1.64048)
        .. controls (-0.36230,1.73656) and (-0.56766,1.81073)
        .. (-0.76247,1.89008);
    \else
      \draw[appendix crossing] (-0.70877,1.57002)
        .. controls (-0.60007,1.71281) and (-0.39539,1.82455)
        .. (-0.21873,1.94314);
    \fi
    \draw[appendix strand] (HTRupper)
      .. controls (0.58,0.75) and (0.25,0.22)
      .. coordinate[pos=0.50] (HTREdge)
         coordinate[pos=0.60] (HTREdgeForward) (HTRlower);
    \draw[appendix strand] (HTRbottom) -- (HTRlower);
    \node[appendix vertex] at (HTRupper) {};
    \node[appendix vertex] at (HTRlower) {};
    \node[appendix label, above=0pt of HTRaTop] {$a$};
    \node[appendix label, above=0pt of HTRbTop] {$b$};
    \node[appendix label, above=0pt of HTRcTop] {$c$};
    \HexagonChannelAt{HTREdge}{HTREdgeForward}{90}{g}
    \node[appendix label, below=0pt of HTRbottom] {$d$};
  \end{tikzpicture}%
}

\newcommand{\HexagonTargetBraidedTree}[1]{%
  \begin{tikzpicture}[baseline=-0.52ex, x=0.48cm, y=0.45cm]
    \path[use as bounding box] (-1.48,-1.10) rectangle (1.48,2.42);
    \coordinate (HTaTop) at (-1.20,2.12);
    \coordinate (HTbTop) at (0,2.12);
    \coordinate (HTcTop) at (1.20,2.12);
    \coordinate (HTupper) at (0.64,1.02);
    \coordinate (HTlower) at (0,0);
    \coordinate (HTbottom) at (0,-0.82);
    \coordinate (HTcTurn) at (0.28,1.32);
    \coordinate (HTaTurnOne) at (0.94,1.34);
    \coordinate (HTaTurnTwo) at (0.28,1.60);
    \coordinate (HTbTurn) at (-0.78,1.45);
    \draw[appendix strand] (HTupper)
      .. controls (0.48,1.18) and (0.30,1.20) .. (HTcTurn)
      .. controls (0.40,1.60) and (0.95,1.80) .. (HTcTop);
    \draw[appendix strand] (HTlower)
      .. controls (-0.45,0.42) and (-0.85,1.05) .. (HTbTurn)
      .. controls (-0.70,1.74) and (-0.25,1.92) .. (HTbTop);
    \draw[appendix strand] (HTupper)
      .. controls (0.82,1.12) and (0.98,1.20) .. (HTaTurnOne)
      .. controls (0.82,1.56) and (0.48,1.56) .. (HTaTurnTwo)
      .. controls (-0.15,1.66) and (-0.78,1.84) .. (HTaTop);
    \ifnum#1=1
      \draw[appendix crossing] (0.81865,1.46518)
        .. controls (0.66296,1.56236) and (0.43138,1.56972) .. (HTaTurnTwo)
        .. controls (0.24228,1.60526) and (0.20302,1.61145)
        .. (0.16250,1.61855);
      \draw[appendix crossing] (-0.20148,1.69926)
        .. controls (-0.39613,1.75106) and (-0.59974,1.81887)
        .. (-0.79051,1.90164);
    \else
      \draw[appendix crossing] (0.28446,1.30185)
        .. controls (0.28255,1.30765) and (0.28105,1.31368) .. (HTcTurn)
        .. controls (0.34947,1.48211) and (0.56307,1.61740)
        .. (0.77915,1.76468);
      \draw[appendix crossing] (-0.72472,1.56793)
        .. controls (-0.62132,1.72309) and (-0.41611,1.84600)
        .. (-0.22853,1.96389);
    \fi
    \draw[appendix strand] (HTupper)
      .. controls (0.57,0.70) and (0.25,0.22)
      .. coordinate[pos=0.50] (HTEdge)
         coordinate[pos=0.60] (HTEdgeForward) (HTlower);
    \draw[appendix strand] (HTbottom) -- (HTlower);
    \node[appendix vertex] at (HTupper) {};
    \node[appendix vertex] at (HTlower) {};
    \node[appendix label, above=0pt of HTaTop] {$a$};
    \node[appendix label, above=0pt of HTbTop] {$b$};
    \node[appendix label, above=0pt of HTcTop] {$c$};
    \HexagonChannelAt{HTEdge}{HTEdgeForward}{90}{g}
    \node[appendix label, below=0pt of HTbottom] {$d$};
  \end{tikzpicture}%
}

\newcommand{\HexagonBottomRightBraidedTree}[1]{%
  \begin{tikzpicture}[baseline=-0.52ex, x=0.48cm, y=0.45cm]
    \path[use as bounding box] (-1.48,-1.10) rectangle (1.48,2.42);
    \coordinate (HBRaTop) at (-1.20,2.12);
    \coordinate (HBRbTop) at (0,2.12);
    \coordinate (HBRcTop) at (1.20,2.12);
    \coordinate (HBRupper) at (0.60,1.18);
    \coordinate (HBRlower) at (0,0);
    \coordinate (HBRbottom) at (0,-0.82);
    \coordinate (HBRfAbove) at (0.42,0.95);
    \coordinate (HBRfBend) at (-0.38,0.48);
    \coordinate (HBRaBend) at (0.38,0.48);
    \coordinate (HBRaAbove) at (-0.45,0.95);
    \draw[appendix strand] (HBRcTop)
      .. controls (1.20,1.72) and (0.88,1.36) .. (HBRupper);
    \draw[appendix strand] (HBRbTop)
      .. controls (0,1.72) and (0.30,1.36) .. (HBRupper);
    \draw[appendix strand] (HBRupper)
      .. controls (0.54,1.08) and (0.48,1.00)
      .. (HBRfAbove)
      .. controls (0.20,0.82) and (-0.38,0.72) .. (HBRfBend)
      .. controls (-0.38,0.28) and (-0.12,0.15) .. (HBRlower);
    \draw[appendix strand] (HBRlower)
      .. controls (0.12,0.15) and (0.38,0.28) .. (HBRaBend)
      .. controls (0.38,0.72) and (-0.20,0.82) .. (HBRaAbove)
      .. controls (-0.66,1.08) and (-1.02,1.72) .. (HBRaTop);
    \ifnum#1=1
      \draw[appendix crossing] (0.31885,0.60741)
        .. controls (0.18660,0.73777) and (-0.13146,0.81871)
        .. (-0.34703,0.90409);
    \else
      \draw[appendix crossing] (0.33751,0.90880)
        .. controls (0.13633,0.82073) and (-0.19225,0.73840)
        .. (-0.32337,0.60308);
    \fi
    \draw[appendix strand] (HBRbottom) -- (HBRlower);
    \node[appendix vertex] at (HBRupper) {};
    \node[appendix vertex] at (HBRlower) {};
    \node[appendix label, above=0pt of HBRaTop] {$a$};
    \node[appendix label, above=0pt of HBRbTop] {$b$};
    \node[appendix label, above=0pt of HBRcTop] {$c$};
    \coordinate (HBRfLabel) at ($(HBRupper)+(180:9pt)$);
    \node[appendix channel] at (HBRfLabel) {$f$};
    \node[appendix label, below=0pt of HBRbottom] {$d$};
  \end{tikzpicture}%
}

\newcommand{\HexagonRadius}{3.20cm}
\newcommand{\HexagonLabelClearance}{0.22cm}

\newcommand{\HexagonEdge}[8]{%
  \coordinate (Hexagon#1#2Start) at
    ($(Hexagon#1.center)!#3!(Hexagon#2.center)$);
  \coordinate (Hexagon#1#2End) at
    ($(Hexagon#2.center)!#4!(Hexagon#1.center)$);
  \coordinate (Hexagon#1#2Mid) at
    ($(Hexagon#1.center)!0.50!(Hexagon#2.center)$);
  \coordinate (Hexagon#1#2LabelAnchor) at
    ($(Hexagon#1#2Mid)+(#5:\HexagonLabelClearance)$);
  \draw[appendix map] (Hexagon#1#2Start) -- (Hexagon#1#2End);
  \node[appendix map label, anchor=#6, yshift=#7]
    at (Hexagon#1#2LabelAnchor) {#8};
}

\newcommand{\HexagonCoherenceDiagramWith}[4]{%
  \begin{tikzpicture}[
    appendix map/.style={
      draw=black,
      line width=0.70pt,
      -{Stealth[length=4.2pt,width=3.3pt]}
    },
    appendix map label/.style={
      font=\small,
      inner sep=0pt
    }
  ]
    \node[inner sep=0pt] (HexagonA) at (180:\HexagonRadius)
      {\HexagonLeftTree{a}{b}{c}{e}{d}};
    \node[inner sep=0pt] (HexagonB) at (120:\HexagonRadius)
      {\HexagonTopLeftBraidedTree{#1}};
    \node[inner sep=0pt] (HexagonC) at (60:\HexagonRadius)
      {\HexagonTopRightBraidedTree{#1}};
    \node[inner sep=0pt] (HexagonD) at (0:\HexagonRadius)
      {\HexagonTargetBraidedTree{#1}};
    \node[inner sep=0pt] (HexagonE) at (-60:\HexagonRadius)
      {\HexagonBottomRightBraidedTree{#1}};
    \node[inner sep=0pt] (HexagonF) at (-120:\HexagonRadius)
      {\HexagonRightTree{a}{b}{c}{f}{d}};

    \HexagonEdge{A}{B}{1.18cm}{1.05cm}{150}{south east}{0pt}{#2}
    \HexagonEdge{B}{C}{1.115cm}{1.115cm}{90}{south}{0pt}{$[F_d^{bac}]_{eg}$}
    \HexagonEdge{C}{D}{1.05cm}{1.18cm}{30}{south west}{0pt}{#3}
    \HexagonEdge{A}{F}{1.05cm}{1.18cm}{-150}{north east}{0pt}{$[F_d^{abc}]_{ef}$}
    \HexagonEdge{F}{E}{1.115cm}{1.115cm}{-90}{north}{0pt}{#4}
    \HexagonEdge{E}{D}{1.18cm}{1.05cm}{-30}{north west}{0pt}{$[F_d^{bca}]_{fg}$}
  \end{tikzpicture}%
}

\newcommand{\HexagonCoherenceDiagram}{%
  \HexagonCoherenceDiagramWith{1}{$R_e^{ba}$}{$R_g^{ca}$}{$R_d^{fa}$}%
}

\newcommand{\InverseHexagonCoherenceDiagram}{%
  \HexagonCoherenceDiagramWith{0}%
    {$(R_e^{ab})^{-1}$}{$(R_g^{ac})^{-1}$}{$(R_d^{af})^{-1}$}%
}

\tikzset{
  appendix strand/.style={
    draw=black,
    line width=0.75pt,
    line cap=round,
    line join=round
  },
  appendix crossing/.style={
    appendix strand,
    preaction={draw=white, line width=3.2pt}
  },
  appendix vertex/.style={circle, fill=black, inner sep=1.25pt},
  appendix label/.style={font=\scriptsize, inner sep=1pt},
  appendix channel/.style={font=\scriptsize, inner sep=0pt},
  appendix basis/.style={font=\scriptsize, inner sep=0pt}
}

\newcommand{\DrawRibbonCenter}[1]{%
    \draw[#1] (0.000000,-2.000000)
      .. controls (0.000000,-1.750000) and (-0.093000,-1.512210) .. (-0.210000,-1.255800)
      .. controls (-0.231667,-1.208317) and (-0.253125,-1.161971) .. (-0.274375,-1.116741)
      .. controls (-0.295625,-1.071510) and (-0.316667,-1.027396) .. (-0.337500,-0.984375)
      .. controls (-0.358333,-0.941354) and (-0.378958,-0.899427) .. (-0.399375,-0.858572)
      .. controls (-0.419792,-0.817717) and (-0.440000,-0.777933) .. (-0.460000,-0.739200)
      .. controls (-0.480000,-0.700467) and (-0.499792,-0.662783) .. (-0.519375,-0.626128)
      .. controls (-0.538958,-0.589473) and (-0.558333,-0.553846) .. (-0.577500,-0.519225)
      .. controls (-0.596667,-0.484604) and (-0.615625,-0.450990) .. (-0.634375,-0.418359)
      .. controls (-0.653125,-0.385729) and (-0.671667,-0.354083) .. (-0.690000,-0.323400)
      .. controls (-0.708333,-0.292717) and (-0.726458,-0.262996) .. (-0.744375,-0.234216)
      .. controls (-0.762292,-0.205435) and (-0.780000,-0.177596) .. (-0.797500,-0.150675)
      .. controls (-0.815000,-0.123754) and (-0.832292,-0.097752) .. (-0.849375,-0.072647)
      .. controls (-0.866458,-0.047542) and (-0.883333,-0.023333) .. (-0.900000,0.000000)
      .. controls (-0.916667,0.023333) and (-0.933125,0.045792) .. (-0.949375,0.067397)
      .. controls (-0.965625,0.089002) and (-0.981667,0.109754) .. (-0.997500,0.129675)
      .. controls (-1.013333,0.149596) and (-1.028958,0.168685) .. (-1.044375,0.186966)
      .. controls (-1.059792,0.205246) and (-1.075000,0.222717) .. (-1.090000,0.239400)
      .. controls (-1.105000,0.256083) and (-1.119792,0.271979) .. (-1.134375,0.287109)
      .. controls (-1.148958,0.302240) and (-1.163333,0.316604) .. (-1.177500,0.330225)
      .. controls (-1.191667,0.343846) and (-1.205625,0.356723) .. (-1.219375,0.368878)
      .. controls (-1.233125,0.381033) and (-1.246667,0.392467) .. (-1.260000,0.403200)
      .. controls (-1.273333,0.413933) and (-1.286458,0.423967) .. (-1.299375,0.433322)
      .. controls (-1.312292,0.442677) and (-1.325000,0.451354) .. (-1.337500,0.459375)
      .. controls (-1.350000,0.467396) and (-1.362292,0.474760) .. (-1.374375,0.481491)
      .. controls (-1.386458,0.488221) and (-1.398333,0.494317) .. (-1.410000,0.499800)
      .. controls (-1.421667,0.505283) and (-1.433125,0.510154) .. (-1.444375,0.514434)
      .. controls (-1.455625,0.518715) and (-1.466667,0.522404) .. (-1.477500,0.525525)
      .. controls (-1.488333,0.528646) and (-1.498958,0.531198) .. (-1.509375,0.533203)
      .. controls (-1.519792,0.535208) and (-1.530000,0.536667) .. (-1.540000,0.537600)
      .. controls (-1.550000,0.538533) and (-1.559792,0.538942) .. (-1.569375,0.538847)
      .. controls (-1.578958,0.538752) and (-1.588333,0.538154) .. (-1.597500,0.537075)
      .. controls (-1.606667,0.535996) and (-1.615625,0.534435) .. (-1.624375,0.532416)
      .. controls (-1.633125,0.530396) and (-1.641667,0.527917) .. (-1.650000,0.525000)
      .. controls (-1.658333,0.522083) and (-1.666458,0.518729) .. (-1.674375,0.514959)
      .. controls (-1.682292,0.511190) and (-1.690000,0.507004) .. (-1.697500,0.502425)
      .. controls (-1.705000,0.497846) and (-1.712292,0.492873) .. (-1.719375,0.487528)
      .. controls (-1.726458,0.482183) and (-1.733333,0.476467) .. (-1.740000,0.470400)
      .. controls (-1.746667,0.464333) and (-1.753125,0.457917) .. (-1.759375,0.451172)
      .. controls (-1.765625,0.444427) and (-1.771667,0.437354) .. (-1.777500,0.429975)
      .. controls (-1.783333,0.422596) and (-1.788958,0.414910) .. (-1.794375,0.406941)
      .. controls (-1.799792,0.398971) and (-1.805000,0.390717) .. (-1.810000,0.382200)
      .. controls (-1.815000,0.373683) and (-1.819792,0.364904) .. (-1.824375,0.355884)
      .. controls (-1.828958,0.346865) and (-1.833333,0.337604) .. (-1.837500,0.328125)
      .. controls (-1.841667,0.318646) and (-1.845625,0.308948) .. (-1.849375,0.299053)
      .. controls (-1.853125,0.289158) and (-1.856667,0.279067) .. (-1.860000,0.268800)
      .. controls (-1.863333,0.258533) and (-1.866458,0.248092) .. (-1.869375,0.237497)
      .. controls (-1.872292,0.226902) and (-1.875000,0.216154) .. (-1.877500,0.205275)
      .. controls (-1.880000,0.194396) and (-1.882292,0.183385) .. (-1.884375,0.172266)
      .. controls (-1.886458,0.161146) and (-1.888333,0.149917) .. (-1.890000,0.138600)
      .. controls (-1.891667,0.127283) and (-1.893125,0.115879) .. (-1.894375,0.104409)
      .. controls (-1.895625,0.092940) and (-1.896667,0.081404) .. (-1.897500,0.069825)
      .. controls (-1.898333,0.058246) and (-1.898958,0.046623) .. (-1.899375,0.034978)
      .. controls (-1.899792,0.023333) and (-1.900000,0.011667) .. (-1.900000,0.000000)
      .. controls (-1.900000,-0.011667) and (-1.899792,-0.023333) .. (-1.899375,-0.034978)
      .. controls (-1.898958,-0.046623) and (-1.898333,-0.058246) .. (-1.897500,-0.069825)
      .. controls (-1.896667,-0.081404) and (-1.895625,-0.092940) .. (-1.894375,-0.104409)
      .. controls (-1.893125,-0.115879) and (-1.891667,-0.127283) .. (-1.890000,-0.138600)
      .. controls (-1.888333,-0.149917) and (-1.886458,-0.161146) .. (-1.884375,-0.172266)
      .. controls (-1.882292,-0.183385) and (-1.880000,-0.194396) .. (-1.877500,-0.205275)
      .. controls (-1.875000,-0.216154) and (-1.872292,-0.226902) .. (-1.869375,-0.237497)
      .. controls (-1.866458,-0.248092) and (-1.863333,-0.258533) .. (-1.860000,-0.268800)
      .. controls (-1.856667,-0.279067) and (-1.853125,-0.289158) .. (-1.849375,-0.299053)
      .. controls (-1.845625,-0.308948) and (-1.841667,-0.318646) .. (-1.837500,-0.328125)
      .. controls (-1.833333,-0.337604) and (-1.828958,-0.346865) .. (-1.824375,-0.355884)
      .. controls (-1.819792,-0.364904) and (-1.815000,-0.373683) .. (-1.810000,-0.382200)
      .. controls (-1.805000,-0.390717) and (-1.799792,-0.398971) .. (-1.794375,-0.406941)
      .. controls (-1.788958,-0.414910) and (-1.783333,-0.422596) .. (-1.777500,-0.429975)
      .. controls (-1.771667,-0.437354) and (-1.765625,-0.444427) .. (-1.759375,-0.451172)
      .. controls (-1.753125,-0.457917) and (-1.746667,-0.464333) .. (-1.740000,-0.470400)
      .. controls (-1.733333,-0.476467) and (-1.726458,-0.482183) .. (-1.719375,-0.487528)
      .. controls (-1.712292,-0.492873) and (-1.705000,-0.497846) .. (-1.697500,-0.502425)
      .. controls (-1.690000,-0.507004) and (-1.682292,-0.511190) .. (-1.674375,-0.514959)
      .. controls (-1.666458,-0.518729) and (-1.658333,-0.522083) .. (-1.650000,-0.525000)
      .. controls (-1.641667,-0.527917) and (-1.633125,-0.530396) .. (-1.624375,-0.532416)
      .. controls (-1.615625,-0.534435) and (-1.606667,-0.535996) .. (-1.597500,-0.537075)
      .. controls (-1.588333,-0.538154) and (-1.578958,-0.538752) .. (-1.569375,-0.538847)
      .. controls (-1.559792,-0.538942) and (-1.550000,-0.538533) .. (-1.540000,-0.537600)
      .. controls (-1.530000,-0.536667) and (-1.519792,-0.535208) .. (-1.509375,-0.533203)
      .. controls (-1.498958,-0.531198) and (-1.488333,-0.528646) .. (-1.477500,-0.525525)
      .. controls (-1.466667,-0.522404) and (-1.455625,-0.518715) .. (-1.444375,-0.514434)
      .. controls (-1.433125,-0.510154) and (-1.421667,-0.505283) .. (-1.410000,-0.499800)
      .. controls (-1.398333,-0.494317) and (-1.386458,-0.488221) .. (-1.374375,-0.481491)
      .. controls (-1.362292,-0.474760) and (-1.350000,-0.467396) .. (-1.337500,-0.459375)
      .. controls (-1.325000,-0.451354) and (-1.312292,-0.442677) .. (-1.299375,-0.433322)
      .. controls (-1.286458,-0.423967) and (-1.273333,-0.413933) .. (-1.260000,-0.403200)
      .. controls (-1.246667,-0.392467) and (-1.233125,-0.381033) .. (-1.219375,-0.368878)
      .. controls (-1.205625,-0.356723) and (-1.191667,-0.343846) .. (-1.177500,-0.330225)
      .. controls (-1.163333,-0.316604) and (-1.148958,-0.302240) .. (-1.134375,-0.287109)
      .. controls (-1.119792,-0.271979) and (-1.105000,-0.256083) .. (-1.090000,-0.239400)
      .. controls (-1.075000,-0.222717) and (-1.059792,-0.205246) .. (-1.044375,-0.186966)
      .. controls (-1.028958,-0.168685) and (-1.013333,-0.149596) .. (-0.997500,-0.129675)
      .. controls (-0.981667,-0.109754) and (-0.965625,-0.089002) .. (-0.949375,-0.067397)
      .. controls (-0.933125,-0.045792) and (-0.916667,-0.023333) .. (-0.900000,0.000000)
      .. controls (-0.883333,0.023333) and (-0.866458,0.047542) .. (-0.849375,0.072647)
      .. controls (-0.832292,0.097752) and (-0.815000,0.123754) .. (-0.797500,0.150675)
      .. controls (-0.780000,0.177596) and (-0.762292,0.205435) .. (-0.744375,0.234216)
      .. controls (-0.726458,0.262996) and (-0.708333,0.292717) .. (-0.690000,0.323400)
      .. controls (-0.671667,0.354083) and (-0.653125,0.385729) .. (-0.634375,0.418359)
      .. controls (-0.615625,0.450990) and (-0.596667,0.484604) .. (-0.577500,0.519225)
      .. controls (-0.558333,0.553846) and (-0.538958,0.589473) .. (-0.519375,0.626128)
      .. controls (-0.499792,0.662783) and (-0.480000,0.700467) .. (-0.460000,0.739200)
      .. controls (-0.440000,0.777933) and (-0.419792,0.817717) .. (-0.399375,0.858572)
      .. controls (-0.378958,0.899427) and (-0.358333,0.941354) .. (-0.337500,0.984375)
      .. controls (-0.316667,1.027396) and (-0.295625,1.071510) .. (-0.274375,1.116741)
      .. controls (-0.253125,1.161971) and (-0.231667,1.208317) .. (-0.210000,1.255800)
      .. controls (-0.093000,1.512210) and (0.000000,1.750000) .. (0.000000,2.000000);
}
\newcommand{\DrawRibbonCenterOverpass}[1]{%
    \draw[#1] (-1.777500,-0.429975)
      .. controls (-1.771667,-0.437354) and (-1.765625,-0.444427) .. (-1.759375,-0.451172)
      .. controls (-1.753125,-0.457917) and (-1.746667,-0.464333) .. (-1.740000,-0.470400)
      .. controls (-1.733333,-0.476467) and (-1.726458,-0.482183) .. (-1.719375,-0.487528)
      .. controls (-1.712292,-0.492873) and (-1.705000,-0.497846) .. (-1.697500,-0.502425)
      .. controls (-1.690000,-0.507004) and (-1.682292,-0.511190) .. (-1.674375,-0.514959)
      .. controls (-1.666458,-0.518729) and (-1.658333,-0.522083) .. (-1.650000,-0.525000)
      .. controls (-1.641667,-0.527917) and (-1.633125,-0.530396) .. (-1.624375,-0.532416)
      .. controls (-1.615625,-0.534435) and (-1.606667,-0.535996) .. (-1.597500,-0.537075)
      .. controls (-1.588333,-0.538154) and (-1.578958,-0.538752) .. (-1.569375,-0.538847)
      .. controls (-1.559792,-0.538942) and (-1.550000,-0.538533) .. (-1.540000,-0.537600)
      .. controls (-1.530000,-0.536667) and (-1.519792,-0.535208) .. (-1.509375,-0.533203)
      .. controls (-1.498958,-0.531198) and (-1.488333,-0.528646) .. (-1.477500,-0.525525)
      .. controls (-1.466667,-0.522404) and (-1.455625,-0.518715) .. (-1.444375,-0.514434)
      .. controls (-1.433125,-0.510154) and (-1.421667,-0.505283) .. (-1.410000,-0.499800)
      .. controls (-1.398333,-0.494317) and (-1.386458,-0.488221) .. (-1.374375,-0.481491)
      .. controls (-1.362292,-0.474760) and (-1.350000,-0.467396) .. (-1.337500,-0.459375)
      .. controls (-1.325000,-0.451354) and (-1.312292,-0.442677) .. (-1.299375,-0.433322)
      .. controls (-1.286458,-0.423967) and (-1.273333,-0.413933) .. (-1.260000,-0.403200)
      .. controls (-1.246667,-0.392467) and (-1.233125,-0.381033) .. (-1.219375,-0.368878)
      .. controls (-1.205625,-0.356723) and (-1.191667,-0.343846) .. (-1.177500,-0.330225)
      .. controls (-1.163333,-0.316604) and (-1.148958,-0.302240) .. (-1.134375,-0.287109)
      .. controls (-1.119792,-0.271979) and (-1.105000,-0.256083) .. (-1.090000,-0.239400)
      .. controls (-1.075000,-0.222717) and (-1.059792,-0.205246) .. (-1.044375,-0.186966)
      .. controls (-1.028958,-0.168685) and (-1.013333,-0.149596) .. (-0.997500,-0.129675)
      .. controls (-0.981667,-0.109754) and (-0.965625,-0.089002) .. (-0.949375,-0.067397)
      .. controls (-0.933125,-0.045792) and (-0.916667,-0.023333) .. (-0.900000,0.000000)
      .. controls (-0.883333,0.023333) and (-0.866458,0.047542) .. (-0.849375,0.072647)
      .. controls (-0.832292,0.097752) and (-0.815000,0.123754) .. (-0.797500,0.150675)
      .. controls (-0.780000,0.177596) and (-0.762292,0.205435) .. (-0.744375,0.234216)
      .. controls (-0.726458,0.262996) and (-0.708333,0.292717) .. (-0.690000,0.323400)
      .. controls (-0.671667,0.354083) and (-0.653125,0.385729) .. (-0.634375,0.418359)
      .. controls (-0.615625,0.450990) and (-0.596667,0.484604) .. (-0.577500,0.519225)
      .. controls (-0.558333,0.553846) and (-0.538958,0.589473) .. (-0.519375,0.626128)
      .. controls (-0.499792,0.662783) and (-0.480000,0.700467) .. (-0.460000,0.739200)
      .. controls (-0.440000,0.777933) and (-0.419792,0.817717) .. (-0.399375,0.858572)
      .. controls (-0.378958,0.899427) and (-0.358333,0.941354) .. (-0.337500,0.984375)
      .. controls (-0.316667,1.027396) and (-0.295625,1.071510) .. (-0.274375,1.116741)
      .. controls (-0.253125,1.161971) and (-0.231667,1.208317) .. (-0.210000,1.255800)
;
}
\newcommand{\DrawRibbonLeft}[1]{%
    \draw[#1] (-0.220000,-2.000000)
      .. controls (-0.220000,-1.750000) and (-0.294929,-1.599635) .. (-0.410148,-1.347128)
      .. controls (-0.431485,-1.300367) and (-0.452601,-1.254761) .. (-0.473494,-1.210291)
      .. controls (-0.494387,-1.165820) and (-0.515058,-1.122483) .. (-0.535505,-1.080261)
      .. controls (-0.555951,-1.038039) and (-0.576174,-0.996931) .. (-0.596170,-0.956917)
      .. controls (-0.616166,-0.916903) and (-0.635937,-0.877982) .. (-0.655479,-0.840136)
      .. controls (-0.675021,-0.802289) and (-0.694335,-0.765516) .. (-0.713418,-0.729797)
      .. controls (-0.732501,-0.694078) and (-0.751354,-0.659413) .. (-0.769973,-0.625781)
      .. controls (-0.788592,-0.592150) and (-0.806977,-0.559552) .. (-0.825126,-0.527969)
      .. controls (-0.843274,-0.496385) and (-0.861186,-0.465816) .. (-0.878856,-0.436242)
      .. controls (-0.896527,-0.406668) and (-0.913957,-0.378088) .. (-0.931141,-0.350484)
      .. controls (-0.948326,-0.322880) and (-0.965265,-0.296252) .. (-0.981953,-0.270579)
      .. controls (-0.998641,-0.244907) and (-1.015079,-0.220192) .. (-1.031259,-0.196413)
      .. controls (-1.047439,-0.172635) and (-1.063363,-0.149795) .. (-1.079021,-0.127872)
      .. controls (-1.094680,-0.105950) and (-1.110074,-0.084947) .. (-1.125195,-0.064843)
      .. controls (-1.140315,-0.044739) and (-1.155163,-0.025535) .. (-1.169726,-0.007212)
      .. controls (-1.184289,0.011111) and (-1.198569,0.028553) .. (-1.212552,0.045133)
      .. controls (-1.226535,0.061713) and (-1.240222,0.077432) .. (-1.253598,0.092309)
      .. controls (-1.266974,0.107186) and (-1.280038,0.121221) .. (-1.292775,0.134435)
      .. controls (-1.305511,0.147649) and (-1.317919,0.160042) .. (-1.329978,0.171636)
      .. controls (-1.342037,0.183231) and (-1.353748,0.194027) .. (-1.365086,0.204050)
      .. controls (-1.376424,0.214073) and (-1.387389,0.223322) .. (-1.397955,0.231828)
      .. controls (-1.408521,0.240333) and (-1.418687,0.248094) .. (-1.428423,0.255146)
      .. controls (-1.438159,0.262198) and (-1.447466,0.268540) .. (-1.456311,0.274216)
      .. controls (-1.465155,0.279891) and (-1.473538,0.284899) .. (-1.481426,0.289293)
      .. controls (-1.489314,0.293686) and (-1.496708,0.297465) .. (-1.503579,0.300695)
      .. controls (-1.510451,0.303924) and (-1.516799,0.306604) .. (-1.522606,0.308814)
      .. controls (-1.528413,0.311023) and (-1.533678,0.312762) .. (-1.538400,0.314122)
      .. controls (-1.543122,0.315483) and (-1.547301,0.316465) .. (-1.550961,0.317169)
      .. controls (-1.554622,0.317874) and (-1.557763,0.318302) .. (-1.560444,0.318552)
      .. controls (-1.563126,0.318802) and (-1.565346,0.318876) .. (-1.567199,0.318858)
      .. controls (-1.569052,0.318839) and (-1.570538,0.318730) .. (-1.571778,0.318584)
      .. controls (-1.573017,0.318438) and (-1.574012,0.318256) .. (-1.574893,0.318053)
      .. controls (-1.575774,0.317849) and (-1.576543,0.317624) .. (-1.577323,0.317351)
      .. controls (-1.578103,0.317078) and (-1.578894,0.316757) .. (-1.579791,0.316330)
      .. controls (-1.580688,0.315902) and (-1.581691,0.315369) .. (-1.582857,0.314657)
      .. controls (-1.584023,0.313945) and (-1.585352,0.313053) .. (-1.586863,0.311913)
      .. controls (-1.588374,0.310773) and (-1.590067,0.309383) .. (-1.591931,0.307687)
      .. controls (-1.593795,0.305991) and (-1.595830,0.303988) .. (-1.598005,0.301640)
      .. controls (-1.600181,0.299292) and (-1.602497,0.296600) .. (-1.604913,0.293543)
      .. controls (-1.607330,0.290486) and (-1.609847,0.287065) .. (-1.612422,0.283276)
      .. controls (-1.614996,0.279487) and (-1.617629,0.275332) .. (-1.620279,0.270818)
      .. controls (-1.622929,0.266304) and (-1.625596,0.261433) .. (-1.628244,0.256222)
      .. controls (-1.630892,0.251011) and (-1.633520,0.245461) .. (-1.636098,0.239597)
      .. controls (-1.638676,0.233732) and (-1.641203,0.227553) .. (-1.643654,0.221087)
      .. controls (-1.646104,0.214622) and (-1.648477,0.207870) .. (-1.650753,0.200863)
      .. controls (-1.653028,0.193855) and (-1.655205,0.186592) .. (-1.657266,0.179105)
      .. controls (-1.659327,0.171618) and (-1.661272,0.163907) .. (-1.663088,0.156004)
      .. controls (-1.664904,0.148101) and (-1.666591,0.140007) .. (-1.668137,0.131753)
      .. controls (-1.669684,0.123499) and (-1.671090,0.115086) .. (-1.672348,0.106545)
      .. controls (-1.673606,0.098005) and (-1.674715,0.089337) .. (-1.675670,0.080574)
      .. controls (-1.676625,0.071812) and (-1.677425,0.062954) .. (-1.678068,0.054033)
      .. controls (-1.678710,0.045112) and (-1.679193,0.036127) .. (-1.679516,0.027111)
      .. controls (-1.679838,0.018095) and (-1.680000,0.009048) .. (-1.680000,0.000000)
      .. controls (-1.680000,-0.009048) and (-1.679838,-0.018095) .. (-1.679516,-0.027111)
      .. controls (-1.679193,-0.036127) and (-1.678710,-0.045112) .. (-1.678068,-0.054033)
      .. controls (-1.677425,-0.062954) and (-1.676625,-0.071812) .. (-1.675670,-0.080574)
      .. controls (-1.674715,-0.089337) and (-1.673606,-0.098005) .. (-1.672348,-0.106545)
      .. controls (-1.671090,-0.115086) and (-1.669684,-0.123499) .. (-1.668137,-0.131753)
      .. controls (-1.666591,-0.140007) and (-1.664904,-0.148101) .. (-1.663088,-0.156004)
      .. controls (-1.661272,-0.163907) and (-1.659327,-0.171618) .. (-1.657266,-0.179105)
      .. controls (-1.655205,-0.186592) and (-1.653028,-0.193855) .. (-1.650753,-0.200863)
      .. controls (-1.648477,-0.207870) and (-1.646104,-0.214622) .. (-1.643654,-0.221087)
      .. controls (-1.641203,-0.227553) and (-1.638676,-0.233732) .. (-1.636098,-0.239597)
      .. controls (-1.633520,-0.245461) and (-1.630892,-0.251011) .. (-1.628244,-0.256222)
      .. controls (-1.625596,-0.261433) and (-1.622929,-0.266304) .. (-1.620279,-0.270818)
      .. controls (-1.617629,-0.275332) and (-1.614996,-0.279487) .. (-1.612422,-0.283276)
      .. controls (-1.609847,-0.287065) and (-1.607330,-0.290486) .. (-1.604913,-0.293543)
      .. controls (-1.602497,-0.296600) and (-1.600181,-0.299292) .. (-1.598005,-0.301640)
      .. controls (-1.595830,-0.303988) and (-1.593795,-0.305991) .. (-1.591931,-0.307687)
      .. controls (-1.590067,-0.309383) and (-1.588374,-0.310773) .. (-1.586863,-0.311913)
      .. controls (-1.585352,-0.313053) and (-1.584023,-0.313945) .. (-1.582857,-0.314657)
      .. controls (-1.581691,-0.315369) and (-1.580688,-0.315902) .. (-1.579791,-0.316330)
      .. controls (-1.578894,-0.316757) and (-1.578103,-0.317078) .. (-1.577323,-0.317351)
      .. controls (-1.576543,-0.317624) and (-1.575774,-0.317849) .. (-1.574893,-0.318053)
      .. controls (-1.574012,-0.318256) and (-1.573017,-0.318438) .. (-1.571778,-0.318584)
      .. controls (-1.570538,-0.318730) and (-1.569052,-0.318839) .. (-1.567199,-0.318858)
      .. controls (-1.565346,-0.318876) and (-1.563126,-0.318802) .. (-1.560444,-0.318552)
      .. controls (-1.557763,-0.318302) and (-1.554622,-0.317874) .. (-1.550961,-0.317169)
      .. controls (-1.547301,-0.316465) and (-1.543122,-0.315483) .. (-1.538400,-0.314122)
      .. controls (-1.533678,-0.312762) and (-1.528413,-0.311023) .. (-1.522606,-0.308814)
      .. controls (-1.516799,-0.306604) and (-1.510451,-0.303924) .. (-1.503579,-0.300695)
      .. controls (-1.496708,-0.297465) and (-1.489314,-0.293686) .. (-1.481426,-0.289293)
      .. controls (-1.473538,-0.284899) and (-1.465155,-0.279891) .. (-1.456311,-0.274216)
      .. controls (-1.447466,-0.268540) and (-1.438159,-0.262198) .. (-1.428423,-0.255146)
      .. controls (-1.418687,-0.248094) and (-1.408521,-0.240333) .. (-1.397955,-0.231828)
      .. controls (-1.387389,-0.223322) and (-1.376424,-0.214073) .. (-1.365086,-0.204050)
      .. controls (-1.353748,-0.194027) and (-1.342037,-0.183231) .. (-1.329978,-0.171636)
      .. controls (-1.317919,-0.160042) and (-1.305511,-0.147649) .. (-1.292775,-0.134435)
      .. controls (-1.280038,-0.121221) and (-1.266974,-0.107186) .. (-1.253598,-0.092309)
      .. controls (-1.240222,-0.077432) and (-1.226535,-0.061713) .. (-1.212552,-0.045133)
      .. controls (-1.198569,-0.028553) and (-1.184289,-0.011111) .. (-1.169726,0.007212)
      .. controls (-1.155163,0.025535) and (-1.140315,0.044739) .. (-1.125195,0.064843)
      .. controls (-1.110074,0.084947) and (-1.094680,0.105950) .. (-1.079021,0.127872)
      .. controls (-1.063363,0.149795) and (-1.047439,0.172635) .. (-1.031259,0.196413)
      .. controls (-1.015079,0.220192) and (-0.998641,0.244907) .. (-0.981953,0.270579)
      .. controls (-0.965265,0.296252) and (-0.948326,0.322880) .. (-0.931141,0.350484)
      .. controls (-0.913957,0.378088) and (-0.896527,0.406668) .. (-0.878856,0.436242)
      .. controls (-0.861186,0.465816) and (-0.843274,0.496385) .. (-0.825126,0.527969)
      .. controls (-0.806977,0.559552) and (-0.788592,0.592150) .. (-0.769973,0.625781)
      .. controls (-0.751354,0.659413) and (-0.732501,0.694078) .. (-0.713418,0.729797)
      .. controls (-0.694335,0.765516) and (-0.675021,0.802289) .. (-0.655479,0.840136)
      .. controls (-0.635937,0.877982) and (-0.616166,0.916903) .. (-0.596170,0.956917)
      .. controls (-0.576174,0.996931) and (-0.555951,1.038039) .. (-0.535505,1.080261)
      .. controls (-0.515058,1.122483) and (-0.494387,1.165820) .. (-0.473494,1.210291)
      .. controls (-0.452601,1.254761) and (-0.431485,1.300367) .. (-0.410148,1.347128)
      .. controls (-0.294929,1.599635) and (-0.220000,1.750000) .. (-0.220000,2.000000);
}
\newcommand{\DrawRibbonLeftOverpass}[1]{%
    \draw[#1] (-1.604913,-0.293543)
      .. controls (-1.602497,-0.296600) and (-1.600181,-0.299292) .. (-1.598005,-0.301640)
      .. controls (-1.595830,-0.303988) and (-1.593795,-0.305991) .. (-1.591931,-0.307687)
      .. controls (-1.590067,-0.309383) and (-1.588374,-0.310773) .. (-1.586863,-0.311913)
      .. controls (-1.585352,-0.313053) and (-1.584023,-0.313945) .. (-1.582857,-0.314657)
      .. controls (-1.581691,-0.315369) and (-1.580688,-0.315902) .. (-1.579791,-0.316330)
      .. controls (-1.578894,-0.316757) and (-1.578103,-0.317078) .. (-1.577323,-0.317351)
      .. controls (-1.576543,-0.317624) and (-1.575774,-0.317849) .. (-1.574893,-0.318053)
      .. controls (-1.574012,-0.318256) and (-1.573017,-0.318438) .. (-1.571778,-0.318584)
      .. controls (-1.570538,-0.318730) and (-1.569052,-0.318839) .. (-1.567199,-0.318858)
      .. controls (-1.565346,-0.318876) and (-1.563126,-0.318802) .. (-1.560444,-0.318552)
      .. controls (-1.557763,-0.318302) and (-1.554622,-0.317874) .. (-1.550961,-0.317169)
      .. controls (-1.547301,-0.316465) and (-1.543122,-0.315483) .. (-1.538400,-0.314122)
      .. controls (-1.533678,-0.312762) and (-1.528413,-0.311023) .. (-1.522606,-0.308814)
      .. controls (-1.516799,-0.306604) and (-1.510451,-0.303924) .. (-1.503579,-0.300695)
      .. controls (-1.496708,-0.297465) and (-1.489314,-0.293686) .. (-1.481426,-0.289293)
      .. controls (-1.473538,-0.284899) and (-1.465155,-0.279891) .. (-1.456311,-0.274216)
      .. controls (-1.447466,-0.268540) and (-1.438159,-0.262198) .. (-1.428423,-0.255146)
      .. controls (-1.418687,-0.248094) and (-1.408521,-0.240333) .. (-1.397955,-0.231828)
      .. controls (-1.387389,-0.223322) and (-1.376424,-0.214073) .. (-1.365086,-0.204050)
      .. controls (-1.353748,-0.194027) and (-1.342037,-0.183231) .. (-1.329978,-0.171636)
      .. controls (-1.317919,-0.160042) and (-1.305511,-0.147649) .. (-1.292775,-0.134435)
      .. controls (-1.280038,-0.121221) and (-1.266974,-0.107186) .. (-1.253598,-0.092309)
      .. controls (-1.240222,-0.077432) and (-1.226535,-0.061713) .. (-1.212552,-0.045133)
      .. controls (-1.198569,-0.028553) and (-1.184289,-0.011111) .. (-1.169726,0.007212)
      .. controls (-1.155163,0.025535) and (-1.140315,0.044739) .. (-1.125195,0.064843)
      .. controls (-1.110074,0.084947) and (-1.094680,0.105950) .. (-1.079021,0.127872)
      .. controls (-1.063363,0.149795) and (-1.047439,0.172635) .. (-1.031259,0.196413)
      .. controls (-1.015079,0.220192) and (-0.998641,0.244907) .. (-0.981953,0.270579)
      .. controls (-0.965265,0.296252) and (-0.948326,0.322880) .. (-0.931141,0.350484)
      .. controls (-0.913957,0.378088) and (-0.896527,0.406668) .. (-0.878856,0.436242)
      .. controls (-0.861186,0.465816) and (-0.843274,0.496385) .. (-0.825126,0.527969)
      .. controls (-0.806977,0.559552) and (-0.788592,0.592150) .. (-0.769973,0.625781)
      .. controls (-0.751354,0.659413) and (-0.732501,0.694078) .. (-0.713418,0.729797)
      .. controls (-0.694335,0.765516) and (-0.675021,0.802289) .. (-0.655479,0.840136)
      .. controls (-0.635937,0.877982) and (-0.616166,0.916903) .. (-0.596170,0.956917)
      .. controls (-0.576174,0.996931) and (-0.555951,1.038039) .. (-0.535505,1.080261)
      .. controls (-0.515058,1.122483) and (-0.494387,1.165820) .. (-0.473494,1.210291)
      .. controls (-0.452601,1.254761) and (-0.431485,1.300367) .. (-0.410148,1.347128)
;
}
\newcommand{\DrawRibbonRight}[1]{%
    \draw[#1] (0.220000,-2.000000)
      .. controls (0.220000,-1.750000) and (0.108929,-1.424785) .. (-0.009852,-1.164472)
      .. controls (-0.031848,-1.116266) and (-0.053649,-1.069180) .. (-0.075256,-1.023191)
      .. controls (-0.096863,-0.977201) and (-0.118275,-0.932308) .. (-0.139495,-0.888489)
      .. controls (-0.160715,-0.844669) and (-0.181743,-0.801923) .. (-0.202580,-0.760227)
      .. controls (-0.223417,-0.718531) and (-0.244063,-0.677884) .. (-0.264521,-0.638264)
      .. controls (-0.284979,-0.598644) and (-0.305248,-0.560050) .. (-0.325332,-0.522459)
      .. controls (-0.345415,-0.484868) and (-0.365313,-0.448279) .. (-0.385027,-0.412669)
      .. controls (-0.404742,-0.377059) and (-0.424273,-0.342427) .. (-0.443624,-0.308750)
      .. controls (-0.462976,-0.275073) and (-0.482148,-0.242350) .. (-0.501144,-0.210558)
      .. controls (-0.520140,-0.178766) and (-0.538960,-0.147904) .. (-0.557609,-0.117947)
      .. controls (-0.576258,-0.087991) and (-0.594735,-0.058940) .. (-0.613047,-0.030771)
      .. controls (-0.631359,-0.002601) and (-0.649505,0.024688) .. (-0.667491,0.051120)
      .. controls (-0.685477,0.077552) and (-0.703304,0.103128) .. (-0.720979,0.127872)
      .. controls (-0.738653,0.152617) and (-0.756176,0.176530) .. (-0.773555,0.199637)
      .. controls (-0.790935,0.222744) and (-0.808171,0.245044) .. (-0.825274,0.266562)
      .. controls (-0.842377,0.288081) and (-0.859348,0.308818) .. (-0.876198,0.328798)
      .. controls (-0.893048,0.348778) and (-0.909778,0.368001) .. (-0.926402,0.386491)
      .. controls (-0.943026,0.404981) and (-0.959545,0.422737) .. (-0.975975,0.439784)
      .. controls (-0.992405,0.456830) and (-1.008748,0.473167) .. (-1.025022,0.488814)
      .. controls (-1.041296,0.504461) and (-1.057502,0.519419) .. (-1.073664,0.533706)
      .. controls (-1.089826,0.547994) and (-1.105944,0.561611) .. (-1.122045,0.574572)
      .. controls (-1.138146,0.587534) and (-1.154230,0.599839) .. (-1.170327,0.611498)
      .. controls (-1.186424,0.623156) and (-1.202534,0.634168) .. (-1.218689,0.644534)
      .. controls (-1.234845,0.654901) and (-1.251046,0.664622) .. (-1.267324,0.673689)
      .. controls (-1.283602,0.682755) and (-1.299959,0.691168) .. (-1.316421,0.698905)
      .. controls (-1.332883,0.706642) and (-1.349451,0.713704) .. (-1.366144,0.720055)
      .. controls (-1.382837,0.726406) and (-1.399655,0.732047) .. (-1.416600,0.736928)
      .. controls (-1.433544,0.741809) and (-1.450616,0.745931) .. (-1.467789,0.749237)
      .. controls (-1.484962,0.752543) and (-1.502237,0.755032) .. (-1.519556,0.756648)
      .. controls (-1.536874,0.758264) and (-1.554237,0.759007) .. (-1.571551,0.758836)
      .. controls (-1.588865,0.758665) and (-1.606129,0.757578) .. (-1.623222,0.755566)
      .. controls (-1.640316,0.753554) and (-1.657238,0.750615) .. (-1.673857,0.746779)
      .. controls (-1.690476,0.742942) and (-1.706790,0.738209) .. (-1.722677,0.732649)
      .. controls (-1.738564,0.727088) and (-1.754023,0.720702) .. (-1.768959,0.713589)
      .. controls (-1.783896,0.706477) and (-1.798309,0.698640) .. (-1.812143,0.690193)
      .. controls (-1.825977,0.681747) and (-1.839231,0.672692) .. (-1.851887,0.663143)
      .. controls (-1.864542,0.653594) and (-1.876600,0.643550) .. (-1.888069,0.633113)
      .. controls (-1.899538,0.622676) and (-1.910420,0.611846) .. (-1.920745,0.600704)
      .. controls (-1.931069,0.589562) and (-1.940837,0.578109) .. (-1.950087,0.566407)
      .. controls (-1.959337,0.554706) and (-1.968070,0.542756) .. (-1.976328,0.530605)
      .. controls (-1.984587,0.518454) and (-1.992371,0.506102) .. (-1.999721,0.493582)
      .. controls (-2.007071,0.481063) and (-2.013987,0.468376) .. (-2.020506,0.455547)
      .. controls (-2.027025,0.442718) and (-2.033146,0.429747) .. (-2.038902,0.416653)
      .. controls (-2.044657,0.403560) and (-2.050047,0.390343) .. (-2.055096,0.377019)
      .. controls (-2.060146,0.363695) and (-2.064856,0.350263) .. (-2.069247,0.336737)
      .. controls (-2.073639,0.323212) and (-2.077712,0.309591) .. (-2.081484,0.295889)
      .. controls (-2.085257,0.282187) and (-2.088728,0.268402) .. (-2.091912,0.254546)
      .. controls (-2.095096,0.240691) and (-2.097992,0.226764) .. (-2.100613,0.212779)
      .. controls (-2.103233,0.198793) and (-2.105577,0.184748) .. (-2.107652,0.170655)
      .. controls (-2.109728,0.156562) and (-2.111535,0.142421) .. (-2.113080,0.128244)
      .. controls (-2.114625,0.114067) and (-2.115908,0.099854) .. (-2.116932,0.085617)
      .. controls (-2.117957,0.071380) and (-2.118724,0.057119) .. (-2.119234,0.042845)
      .. controls (-2.119745,0.028571) and (-2.120000,0.014286) .. (-2.120000,0.000000)
      .. controls (-2.120000,-0.014286) and (-2.119745,-0.028571) .. (-2.119234,-0.042845)
      .. controls (-2.118724,-0.057119) and (-2.117957,-0.071380) .. (-2.116932,-0.085617)
      .. controls (-2.115908,-0.099854) and (-2.114625,-0.114067) .. (-2.113080,-0.128244)
      .. controls (-2.111535,-0.142421) and (-2.109728,-0.156562) .. (-2.107652,-0.170655)
      .. controls (-2.105577,-0.184748) and (-2.103233,-0.198793) .. (-2.100613,-0.212779)
      .. controls (-2.097992,-0.226764) and (-2.095096,-0.240691) .. (-2.091912,-0.254546)
      .. controls (-2.088728,-0.268402) and (-2.085257,-0.282187) .. (-2.081484,-0.295889)
      .. controls (-2.077712,-0.309591) and (-2.073639,-0.323212) .. (-2.069247,-0.336737)
      .. controls (-2.064856,-0.350263) and (-2.060146,-0.363695) .. (-2.055096,-0.377019)
      .. controls (-2.050047,-0.390343) and (-2.044657,-0.403560) .. (-2.038902,-0.416653)
      .. controls (-2.033146,-0.429747) and (-2.027025,-0.442718) .. (-2.020506,-0.455547)
      .. controls (-2.013987,-0.468376) and (-2.007071,-0.481063) .. (-1.999721,-0.493582)
      .. controls (-1.992371,-0.506102) and (-1.984587,-0.518454) .. (-1.976328,-0.530605)
      .. controls (-1.968070,-0.542756) and (-1.959337,-0.554706) .. (-1.950087,-0.566407)
      .. controls (-1.940837,-0.578109) and (-1.931069,-0.589562) .. (-1.920745,-0.600704)
      .. controls (-1.910420,-0.611846) and (-1.899538,-0.622676) .. (-1.888069,-0.633113)
      .. controls (-1.876600,-0.643550) and (-1.864542,-0.653594) .. (-1.851887,-0.663143)
      .. controls (-1.839231,-0.672692) and (-1.825977,-0.681747) .. (-1.812143,-0.690193)
      .. controls (-1.798309,-0.698640) and (-1.783896,-0.706477) .. (-1.768959,-0.713589)
      .. controls (-1.754023,-0.720702) and (-1.738564,-0.727088) .. (-1.722677,-0.732649)
      .. controls (-1.706790,-0.738209) and (-1.690476,-0.742942) .. (-1.673857,-0.746779)
      .. controls (-1.657238,-0.750615) and (-1.640316,-0.753554) .. (-1.623222,-0.755566)
      .. controls (-1.606129,-0.757578) and (-1.588865,-0.758665) .. (-1.571551,-0.758836)
      .. controls (-1.554237,-0.759007) and (-1.536874,-0.758264) .. (-1.519556,-0.756648)
      .. controls (-1.502237,-0.755032) and (-1.484962,-0.752543) .. (-1.467789,-0.749237)
      .. controls (-1.450616,-0.745931) and (-1.433544,-0.741809) .. (-1.416600,-0.736928)
      .. controls (-1.399655,-0.732047) and (-1.382837,-0.726406) .. (-1.366144,-0.720055)
      .. controls (-1.349451,-0.713704) and (-1.332883,-0.706642) .. (-1.316421,-0.698905)
      .. controls (-1.299959,-0.691168) and (-1.283602,-0.682755) .. (-1.267324,-0.673689)
      .. controls (-1.251046,-0.664622) and (-1.234845,-0.654901) .. (-1.218689,-0.644534)
      .. controls (-1.202534,-0.634168) and (-1.186424,-0.623156) .. (-1.170327,-0.611498)
      .. controls (-1.154230,-0.599839) and (-1.138146,-0.587534) .. (-1.122045,-0.574572)
      .. controls (-1.105944,-0.561611) and (-1.089826,-0.547994) .. (-1.073664,-0.533706)
      .. controls (-1.057502,-0.519419) and (-1.041296,-0.504461) .. (-1.025022,-0.488814)
      .. controls (-1.008748,-0.473167) and (-0.992405,-0.456830) .. (-0.975975,-0.439784)
      .. controls (-0.959545,-0.422737) and (-0.943026,-0.404981) .. (-0.926402,-0.386491)
      .. controls (-0.909778,-0.368001) and (-0.893048,-0.348778) .. (-0.876198,-0.328798)
      .. controls (-0.859348,-0.308818) and (-0.842377,-0.288081) .. (-0.825274,-0.266562)
      .. controls (-0.808171,-0.245044) and (-0.790935,-0.222744) .. (-0.773555,-0.199637)
      .. controls (-0.756176,-0.176530) and (-0.738653,-0.152617) .. (-0.720979,-0.127872)
      .. controls (-0.703304,-0.103128) and (-0.685477,-0.077552) .. (-0.667491,-0.051120)
      .. controls (-0.649505,-0.024688) and (-0.631359,0.002601) .. (-0.613047,0.030771)
      .. controls (-0.594735,0.058940) and (-0.576258,0.087991) .. (-0.557609,0.117947)
      .. controls (-0.538960,0.147904) and (-0.520140,0.178766) .. (-0.501144,0.210558)
      .. controls (-0.482148,0.242350) and (-0.462976,0.275073) .. (-0.443624,0.308750)
      .. controls (-0.424273,0.342427) and (-0.404742,0.377059) .. (-0.385027,0.412669)
      .. controls (-0.365313,0.448279) and (-0.345415,0.484868) .. (-0.325332,0.522459)
      .. controls (-0.305248,0.560050) and (-0.284979,0.598644) .. (-0.264521,0.638264)
      .. controls (-0.244063,0.677884) and (-0.223417,0.718531) .. (-0.202580,0.760227)
      .. controls (-0.181743,0.801923) and (-0.160715,0.844669) .. (-0.139495,0.888489)
      .. controls (-0.118275,0.932308) and (-0.096863,0.977201) .. (-0.075256,1.023191)
      .. controls (-0.053649,1.069180) and (-0.031848,1.116266) .. (-0.009852,1.164472)
      .. controls (0.108929,1.424785) and (0.220000,1.750000) .. (0.220000,2.000000);
}
\newcommand{\DrawRibbonRightOverpass}[1]{%
    \draw[#1] (-1.950087,-0.566407)
      .. controls (-1.940837,-0.578109) and (-1.931069,-0.589562) .. (-1.920745,-0.600704)
      .. controls (-1.910420,-0.611846) and (-1.899538,-0.622676) .. (-1.888069,-0.633113)
      .. controls (-1.876600,-0.643550) and (-1.864542,-0.653594) .. (-1.851887,-0.663143)
      .. controls (-1.839231,-0.672692) and (-1.825977,-0.681747) .. (-1.812143,-0.690193)
      .. controls (-1.798309,-0.698640) and (-1.783896,-0.706477) .. (-1.768959,-0.713589)
      .. controls (-1.754023,-0.720702) and (-1.738564,-0.727088) .. (-1.722677,-0.732649)
      .. controls (-1.706790,-0.738209) and (-1.690476,-0.742942) .. (-1.673857,-0.746779)
      .. controls (-1.657238,-0.750615) and (-1.640316,-0.753554) .. (-1.623222,-0.755566)
      .. controls (-1.606129,-0.757578) and (-1.588865,-0.758665) .. (-1.571551,-0.758836)
      .. controls (-1.554237,-0.759007) and (-1.536874,-0.758264) .. (-1.519556,-0.756648)
      .. controls (-1.502237,-0.755032) and (-1.484962,-0.752543) .. (-1.467789,-0.749237)
      .. controls (-1.450616,-0.745931) and (-1.433544,-0.741809) .. (-1.416600,-0.736928)
      .. controls (-1.399655,-0.732047) and (-1.382837,-0.726406) .. (-1.366144,-0.720055)
      .. controls (-1.349451,-0.713704) and (-1.332883,-0.706642) .. (-1.316421,-0.698905)
      .. controls (-1.299959,-0.691168) and (-1.283602,-0.682755) .. (-1.267324,-0.673689)
      .. controls (-1.251046,-0.664622) and (-1.234845,-0.654901) .. (-1.218689,-0.644534)
      .. controls (-1.202534,-0.634168) and (-1.186424,-0.623156) .. (-1.170327,-0.611498)
      .. controls (-1.154230,-0.599839) and (-1.138146,-0.587534) .. (-1.122045,-0.574572)
      .. controls (-1.105944,-0.561611) and (-1.089826,-0.547994) .. (-1.073664,-0.533706)
      .. controls (-1.057502,-0.519419) and (-1.041296,-0.504461) .. (-1.025022,-0.488814)
      .. controls (-1.008748,-0.473167) and (-0.992405,-0.456830) .. (-0.975975,-0.439784)
      .. controls (-0.959545,-0.422737) and (-0.943026,-0.404981) .. (-0.926402,-0.386491)
      .. controls (-0.909778,-0.368001) and (-0.893048,-0.348778) .. (-0.876198,-0.328798)
      .. controls (-0.859348,-0.308818) and (-0.842377,-0.288081) .. (-0.825274,-0.266562)
      .. controls (-0.808171,-0.245044) and (-0.790935,-0.222744) .. (-0.773555,-0.199637)
      .. controls (-0.756176,-0.176530) and (-0.738653,-0.152617) .. (-0.720979,-0.127872)
      .. controls (-0.703304,-0.103128) and (-0.685477,-0.077552) .. (-0.667491,-0.051120)
      .. controls (-0.649505,-0.024688) and (-0.631359,0.002601) .. (-0.613047,0.030771)
      .. controls (-0.594735,0.058940) and (-0.576258,0.087991) .. (-0.557609,0.117947)
      .. controls (-0.538960,0.147904) and (-0.520140,0.178766) .. (-0.501144,0.210558)
      .. controls (-0.482148,0.242350) and (-0.462976,0.275073) .. (-0.443624,0.308750)
      .. controls (-0.424273,0.342427) and (-0.404742,0.377059) .. (-0.385027,0.412669)
      .. controls (-0.365313,0.448279) and (-0.345415,0.484868) .. (-0.325332,0.522459)
      .. controls (-0.305248,0.560050) and (-0.284979,0.598644) .. (-0.264521,0.638264)
      .. controls (-0.244063,0.677884) and (-0.223417,0.718531) .. (-0.202580,0.760227)
      .. controls (-0.181743,0.801923) and (-0.160715,0.844669) .. (-0.139495,0.888489)
      .. controls (-0.118275,0.932308) and (-0.096863,0.977201) .. (-0.075256,1.023191)
      .. controls (-0.053649,1.069180) and (-0.031848,1.116266) .. (-0.009852,1.164472)
;
}

\newcommand{\SimultaneouslyTwistedPair}[1][0]{%
  \begin{tikzpicture}[baseline=-.65ex,x=.55cm,y=.55cm]
    \path[use as bounding box] (-2.25,-3) rectangle (.55,2.4);
    \begin{scope}[shift={(0,.367816092)},yscale=.816091954]
    \DrawRibbonLeft{appendix strand}
    \DrawRibbonRight{appendix strand}
    \DrawRibbonLeftOverpass{appendix crossing}
    \DrawRibbonRightOverpass{appendix crossing}
    \node[appendix label,above=2pt] at (-.30,2) {$a$};
    \node[appendix label,above=2pt] at (.30,2) {$b$};
    \ifnum#1=1
      \draw[appendix strand] (0,-3.697887324) -- (0,-2.35);
      \draw[appendix strand] (0,-2.35)
        .. controls (-.15,-2.23) and (-.22,-2.2) .. (-.22,-2);
      \draw[appendix strand] (0,-2.35)
        .. controls (.15,-2.23) and (.22,-2.2) .. (.22,-2);
      \node[appendix vertex] at (0,-2.35) {};
      \node[appendix label,below=2pt] at (0,-3.697887324) {$c$};
    \else
      \node[appendix label,below=3pt] at (-.22,-2) {$a$};
      \node[appendix label,below=3pt] at (.22,-2) {$b$};
    \fi
    \end{scope}
  \end{tikzpicture}%
}

\newcommand{\DrawBalancingBraid}{%
  \draw[appendix strand] (-.65,-.1)
    .. controls (-.65,.2) and (.65,.4) .. (.65,.7)
    .. controls (.65,1) and (-.65,1.2) .. (-.65,1.5) -- (-.65,2);
  \draw[appendix strand] (.65,-.1)
    .. controls (.65,.2) and (-.65,.4) .. (-.65,.7)
    .. controls (-.65,1) and (.65,1.2) .. (.65,1.5) -- (.65,2);
  \draw[appendix crossing] (-.3358375,.16383)
    .. controls (-.1321625,.25617) and (.1321625,.34383) .. (.3358375,.43617);
  \draw[appendix crossing] (-.3358375,.96383)
    .. controls (-.1321625,1.05617) and (.1321625,1.14383) .. (.3358375,1.23617);
}
\newcommand{\BalancingChannelRoot}{%
  \draw[appendix strand] (0,-3.697887324) -- (0,-2.35);
  \draw[appendix strand] (0,-2.35)
    .. controls (-.25,-2.28) and (-.65,-2.3) .. (-.65,-2.1);
  \draw[appendix strand] (0,-2.35)
    .. controls (.25,-2.28) and (.65,-2.3) .. (.65,-2.1);
  \node[appendix vertex] at (0,-2.35) {};
  \node[appendix label,below=2pt] at (0,-3.697887324) {$c$};
}
\newcommand{\BalancingTopLabels}{%
  \node[appendix label,above=2pt] at (-.65,2) {$a$};
  \node[appendix label,above=2pt] at (.65,2) {$b$};
}
\newcommand{\DoubleBraidWithIndividualTwists}[1][0]{%
  \begin{tikzpicture}[baseline=-.65ex,x=.55cm,y=.55cm]
    \path[use as bounding box] (-1.7,-3) rectangle (.85,2.4);
    \begin{scope}[shift={(0,.367816092)},yscale=.816091954]
    \foreach \x in {-.65,.65}{
      \begin{scope}[shift={(\x,-1.1)},scale=.5]
        \DrawRibbonCenter{appendix strand}
        \DrawRibbonCenterOverpass{appendix crossing}
      \end{scope}
    }
    \DrawBalancingBraid
    \BalancingTopLabels
    \ifnum#1=1
      \BalancingChannelRoot
    \else
      \node[appendix label,below=3pt] at (-.65,-2.1) {$a$};
      \node[appendix label,below=3pt] at (.65,-2.1) {$b$};
    \fi
    \end{scope}
  \end{tikzpicture}%
}

\begin{document}
\hypersetup{citecolor=blue,linkcolor=blue,urlcolor=blue}
\interfootnotelinepenalty=10000
\baselineskip=18pt

\vspace*{2cm}
\thispagestyle{empty}

\begin{center}
{\fontsize{20.74pt}{0pt}  \bf
Computing $F$-Symbols from Modular Data
}

\bigskip\vspace{1cm}{
{\large Jonah Berean-Dutcher$^{\,1,\,2}$, Clay C\'ordova$^{\,3}$, Julio Parra-Martinez$^{\,2}$, and Richard Tuddenham$^{\,3}$}
} \\[7mm]
{$^{1}$\textit{University of British Columbia, Vancouver, V6T 1Z1, Canada} \\
$^{2}$\textit{Institut des Hautes \'Etudes Scientifiques, 91440 Bures-sur-Yvette, France} \\
$^{3}$\textit{Leinweber Institute for Theoretical Physics  \& Enrico Fermi Institute, University of Chicago}} \\
\end{center}

\bigskip
\centerline{\large\bf Abstract}
\begin{quote} \small
We present a strategy for computing the $F$-symbols of unitary modular categories with a self-dual simple generator, starting from their modular data, which almost entirely avoids the non-linear solution of the pentagon equations.
The algorithm climbs a ladder of braid group representations on successively more strands of the generator until enough $F$-symbols have been determined to propagate them to a complete solution. The strategy presented is successful in difficult cases through ranks 11 and 12, including $\mathcal Z(\text{Haagerup}_3)$, and does not require the category to be multiplicity-free. Although the algorithm is implemented numerically, we describe a reliable means of recovering the exact $F$-symbols, assuming that in some gauge they lie in a given number field, and of certifying every pentagon and hexagon equation exactly. The exotic rank-11 anisotropic modular category \href{https://anyonatlas.org/data/B11-00119}{\texttt{B11-00119}} with largest fusion coefficient 4 is used as a motivating example, for which the solve takes roughly 10 minutes on a laptop. We will continue to add $F$-symbol solutions to \href{https://anyonatlas.org}{\texttt{AnyonAtlas.org}} as they are completed.
\end{quote}
\setcounter{footnote}{0}

\vfill
{\small \textit{E-mail:} \href{mailto:jbd@phas.ubc.ca}{\texttt{jbd@phas.ubc.ca}}, \href{mailto:clayc@uchicago.edu}{\texttt{clayc@uchicago.edu}}, \href{mailto:julio@ihes.fr}{\texttt{julio@ihes.fr}}, \href{mailto:rtuddenh@uchicago.edu}{\texttt{rtuddenh@uchicago.edu}}}\par\medskip
\today
\newpage
\tableofcontents

\flushbottom

\section{Introduction}
\label{sec:intro}

The classification of topological quantum field theories (TQFTs) in $2+1$ dimensions is closely tied to the classification of modular tensor categories (MTCs), which encode the fusion and braiding of anyons in bosonic topological phases~\cite{kitaevAnyonsExactlySolved2006,turaevModularCategories3Manifold1992}. Much of the systematic progress in this direction has come from classifying the modular data $(S,T)$~\cite{Rowell:2007dge,ngClassificationModularData2025,Berean-Dutcher:2026fck}, where $S$ is the normalized Hopf-link matrix, encoding mutual braiding, and $T$ records the topological spins. These matrices determine the fusion rules through the Verlinde formula and, in a suitable gauge, also fix the $R$-symbols describing anyon exchange~\cite{ngRecoveringRsymbolsModular2024}. Modular data do not, however, determine the full category: inequivalent modular categories can share the same $(S,T)$~\cite{mignardModularCategoriesAre2021}, and the known constraints on modular data alone do not establish that a candidate admits a categorification.

To construct a category from modular data, one must determine compatible $F$-symbols, which encode the associativity of fusion in chosen bases of the junction spaces. Together with the fusion rules $N^{ab}_c$, these symbols determine a fusion category up to monoidal equivalence; specifying the braiding and twists through $(R,T)$ completes the data of a ribbon fusion category~\cite{kitaevAnyonsExactlySolved2006,bondersonNonAbelianAnyons2007,etingofTensorCategories2015}. Compatibility requires the $F$- and $R$-symbols to satisfy the pentagon and hexagon equations. The main computational obstacle is the large, coupled system of non-linear pentagon equations, whose direct solution rapidly becomes impractical as the rank and fusion multiplicities grow.

\begin{figure}[!htb]
    \centering
    \includegraphics[width=0.57\linewidth]{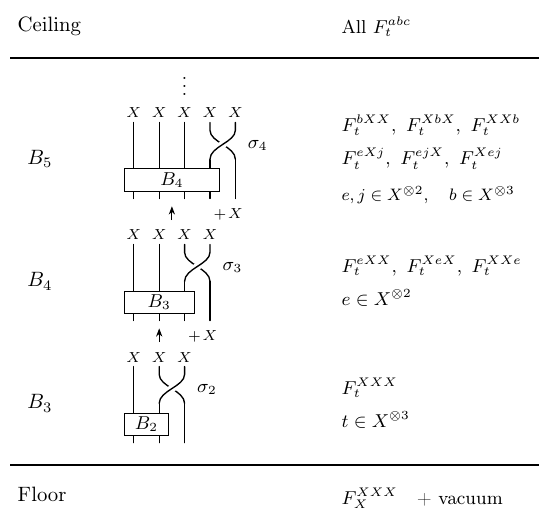}
    \caption{An illustration of the braid group ladder. Here $a\in X^{\otimes n}$ means that $a$ is a simple summand of $X^{\otimes n}$. Boxes denote inherited braid actions; the $F$-matrix families include the pentagon completions at each rung.}
    \label{fig:braid-group-ladder}
\end{figure}

In this paper we present an algorithm to compute compatible $F$-symbols from modular data for unitary modular categories with a self-dual simple generator $X$. Here $X$ is isomorphic to its dual, and every simple object occurs in some tensor power $X^{\otimes n}$. The algorithm uses representations of the braid group $B_n$ on successively more $X$ strands. Each added strand carries forward the previous braid action, leaving one new generator to determine. We call this construction the \emph{braid group ladder} (\cref{fig:braid-group-ladder}). With the $R$-symbols fixed in a diagonal gauge, the bottom of the ladder is a small subsystem centered on $F_X^{XXX}$. Adding successive strands enlarges the solved family of $F$-symbols using hexagons, braid relations, and selected pentagons. Once enough $F$-symbols are known, linear pentagon and hexagon systems propagate this seed to all remaining $F$-symbols. In the cases tested, this almost completely avoids non-linear pentagon solves.

We have implemented the algorithm as a numerical search, which does not guarantee that all solutions are found. When a solution is found, however, we reconstruct the $F$-symbols exactly, under the assumption that they admit a compatible gauge in a chosen number field. We then certify every pentagon and hexagon equation exactly, establishing the validity of the recovered category. The field choice and reconstruction assumptions are discussed in \cref{sec:assumptions,sec:reconstruction}.

Our immediate motivation was to categorify the rank-11 completely anisotropic modular data \href{https://anyonatlas.org/data/B11-00119}{\texttt{B11-00119}}, first identified in~\cite{ngClassificationModularData2025} and proposed in~\cite{Berean-Dutcher:2026fck} as a likely candidate for the first known modular category not Witt-equivalent to any Chern--Simons theory.\footnote{A modular category is completely anisotropic if it admits no non-trivial anyon condensation, equivalently no connected commutative separable algebra other than the tensor unit. Two modular categories $\mathcal C$ and $\mathcal D$ are Witt-equivalent if $\mathcal C\boxtimes\mathcal D^{\mathrm{rev}}$ is braided-equivalent to a Drinfeld center; here $\boxtimes$ denotes the tensor product of categories and $\mathrm{rev}$ reverses the braiding. A Witt class is an equivalence class under this relation~\cite{davydovWittGroupNondegenerate2011}.} We establish a unitary categorification by computing exact, unitary forms of all 3,855,922 $F$-symbols in roughly ten minutes on an M1 chip. This category has the most $F$-symbols of any modular data at its rank~\cite{Berean-Dutcher:2026fck}.

The method also computes exact $F$-symbols of the Drinfeld center $\mathcal Z(\mathcal H_3)$~\cite{hongEXOTICMODULARTENSOR2008,evansExoticnessRealisabilityTwisted2011} of the Haagerup fusion category $\mathcal H_3$, listed as \href{https://anyonatlas.org/data/B12-00071}{\texttt{B12-00071}}, and of the exotic anisotropic rank-12 category \href{https://anyonatlas.org/data/B12-00057}{\texttt{B12-00057}}, both with comparable runtimes. The Haagerup center was previously solved by more tailored approaches in~\cite{maurerFsymbolsRsymbolsDrinfeld2026}.

Beyond these examples, explicit $F$-symbols support both classification and physical applications. Paired with the algorithm for identifying modular data in~\cite{Berean-Dutcher:2026fck}, our method provides a way to construct broad classes of unitary modular categories rank by rank. The $F$-symbols are also the input to string-net lattice models and the associated Turaev--Viro state sums~\cite{levinStringnetCondensationPhysical2005,kirillovStringnetModelTuraevViro2011}, and, together with the $R$-symbols, they determine the braiding gates used in topological quantum computation~\cite{Freedman:2001eqc,Nayak:2008zza,rowellMathematicsTopologicalQuantum2018}.

Previous work on computing the $F$-symbols of fusion categories algorithmically has centered primarily on multiplicity-free cases. Algebraic approaches have followed the general strategy of decomposing the pentagon and hexagon equations into successively solved subsystems~\cite{bondersonNonAbelianAnyons2007,aboumradFMatrixBraidCalculator2022,vercleyen2024lowrankmultiplicityfreefusioncategories,vercleyenLowRankMultiplicityFreeComplex2026}. Vercleyen developed the general software package Anyonica for computing the $F$-symbols of multiplicity-free fusion categories~\cite{vercleyenAnyonica2024}, which was used for a census of multiplicity-free pivotal fusion categories up to rank 7 on \href{https://anyonwiki.org/}{\texttt{AnyonWiki}}~\cite{vercleyenLowRankMultiplicityFreeComplex2026,anyonwikiorg2026}. Unlike our method, Anyonica does not require modularity, unitarity, or a self-dual simple generator $X$. Our numerical approach scales better than these exact methods, enabling computations at higher ranks and with general fusion multiplicities. Other large computations have instead exploited additional structure, such as Morita equivalence to a category with known $F$-symbols~\cite{barterComputingAssociatorsEndomorphism2022} or a description as the Drinfeld center of a known fusion category~\cite{maurerFsymbolsRsymbolsDrinfeld2026}.

Braid group representations can, in some cases, be determined inductively from representations on fewer strands~\cite{martirosyanWenzlBraidRigidity2022,martirosyanWenzlReconstructionG22025,martirosyanWenzlSimpleB42025}. Separately, in work on knot invariants, it was proposed to express the $F$-symbols, known in that context as Racah matrices, in terms of braiding eigenvalues~\cite{itoyamaEigenvalueHypothesis2013}. Later work reused three-strand solutions to obtain the new $F$-symbols appearing in four-strand representations~\cite{dharaEigenvalueHypothesis2018}. This provides a computational precedent for the braid group ladder described here.

This paper is organized as follows. We review the definitions of, and constraints on, the $F$- and $R$-symbols in \cref{sec:prelim}. There, we highlight the gauge fixing of the $R$-symbols in terms of the modular data and describe the braid group representations on $\Hom(t,X^{\otimes n})$ for simple $t$. The strategy itself is described in \cref{sec:alg}, together with the exact reconstruction and certification procedure. Throughout, the exotic rank-11 category of \cref{sec:rank-eleven} serves as a running example. We close in \cref{sec:disc} with the scope of the method and possible extensions.

\section{Preliminaries}
\label{sec:prelim}

We first describe the junction spaces and consistency equations, then explain how modular data fix the braiding and how the resulting braid group representations organize our algorithm. We conclude with the rank-11 example.

\subsection{Skeletal Data of Modular Categories}
\label{sec:skeletal-data}

To work with explicit matrices, we choose representatives of the isomorphism classes of objects and bases of the junction spaces. Every fusion category is equivalent to a skeletal one, in which isomorphic objects are equal; the resulting fusion rules and matrices are called its skeletal data. General accounts can be found in~\cite{kitaevAnyonsExactlySolved2006,bakalovLecturesTensorCategories2000,etingofTensorCategories2015,bondersonNonAbelianAnyons2007}.

A modular category $\mathcal{C}$ has finitely many isomorphism classes of simple objects, representing the anyon types. Their fusion decomposes as
\begin{align}
a \otimes b \cong \bigoplus_{c \in \operatorname{Irr}(\mathcal{C})} N^{ab}_c c\,,~~~~ N^{ab}_c\in\mathbb{Z}_{\geq 0}\label{eq:fusion}
\end{align}
Here $\operatorname{Irr}(\mathcal{C})$ denotes the set of isomorphism classes of simple objects; its cardinality is the rank of $\mathcal C$. We denote the tensor unit (or vacuum) by $1$ and the dual of $a$ by $\bar a$.

The coefficient $N^{ab}_c$ counts the independent junctions splitting $c$ into $a$ and $b$. These junctions form a vector space,
\begin{align}
    V^{ab}_{c}=\Hom_{\C}(c,a\otimes b),~~~~
    N^{ab}_{c} = \dim V^{ab}_c
    \label{eq:splitting-spaces}
\end{align}

A basis of $V^{ab}_c$ therefore labels the independent ways to realize such a junction. The fusion rules are multiplicity-free if every $N^{ab}_c$ is at most one; otherwise, multiplicity indices are needed to distinguish these basis elements.

We write $d_a$ for the quantum dimension of $a$, the categorical trace of $\id_a$ in the chosen spherical structure. In the unitary case, these dimensions are positive. Below, they enter both the normalization of the modular data and constraints on individual $F$-symbols.

For three outgoing anyons, a junction can be resolved into a tree by first grouping $a,b$ or by first grouping $b,c$. The associator $(a\otimes b)\otimes c\cong a\otimes(b\otimes c)$ relates these two groupings and induces the linear isomorphism
\begin{align}
    F^{abc}_d:
    \bigoplus_{e\in\operatorname{Irr}(\mathcal C)}V_d^{ec}\otimes V_e^{ab}
    &\longrightarrow
    \bigoplus_{f\in\operatorname{Irr}(\mathcal C)}V_d^{af}\otimes V_f^{bc}
    \label{eq:f-map}
\end{align}
The sums run over the allowed intermediate simple objects: $e$ for the first grouping and $f$ for the second. Choosing bases for the junctions at each vertex gives two fusion-tree bases. We call the matrix relating them the $F$-matrix $F^{abc}_d$, and its entries the $F$-symbols, defined diagrammatically by
\begin{align}
    \vcenter{\hbox{\LeftAssociatedFusionTree}}
    &=\sum_{f,\mu,\nu}
      [F^{abc}_d]_{(e,\alpha,\beta),(f,\mu,\nu)}
      \vcenter{\hbox{\RightAssociatedFusionTree}}
    \label{eq:f-move}
\end{align}
Here Greek indices label the chosen junction basis elements, e.g.\ $\alpha \in V^{ec}_d$; a circle $\circ$ denotes the sole basis label in a one-dimensional junction space. We write the source fusion-tree index first and the target index second, as in the displayed expansion.

Changing the bases of the junction spaces changes the $F$-symbols without changing the category. This freedom is called gauge redundancy. In a unitary category, orthonormal junction bases give a \emph{unitary gauge}, in which the associator and braiding matrices are unitary. In particular,
\begin{equation}
    F^{abc}_d (F^{abc}_d)^\dagger  = I
\end{equation}

The associator must be consistent when applied repeatedly. For four strands, the pentagon equations require that two different sequences of reassociations between the same initial and final trees give the same map:
\begin{align}
    \vcenter{\hbox{\PentagonCoherenceDiagram}}
    \label{eq:pentagon}
\end{align}
The two sides contain respectively two and three $F$-moves. In components, their equality reads
\begin{multline}
    \sum_{\lambda}
[F_x^{ecd}]_{(f,\alpha,\beta),(g,\lambda,\rho)}
[F_x^{abg}]_{(e,\lambda,\gamma),(h,\mu,\nu)}
=\\
\sum_{k,\sigma,\tau,\kappa}
[F_f^{abc}]_{(e,\beta,\gamma),(k,\sigma,\tau)}
[F_x^{akd}]_{(f,\alpha,\sigma),(h,\mu,\kappa)}
[F_h^{bcd}]_{(k,\kappa,\tau),(g,\nu,\rho)}
\end{multline}
We denote the collection of pentagon equations with outgoing $a,b,c,d$ and incoming $x$ by $P(a,b,c,d;x)$. By Ocneanu rigidity~\cite{etingofFusionCategories2017}, the full system admits no continuous deformations of a fusion category other than gauge transformations, and there are only finitely many solutions up to gauge for fixed fusion rules. This distinction between changes of basis and genuine deformations will matter when we solve smaller, potentially underdetermined subsystems.

A second important condition is the triangle equation, which enforces compatibility between the associator and the unit object $1$. In terms of the $F$-symbols, it requires that $F^{abc}_d \propto I$ if $1\in\{a,b,c\}$. More strongly, there exists a gauge where
\begin{equation}\label{eq:unit-gauge}
    F^{abc}_d = I \quad \text{if } 1\in\{a,b,c\}
\end{equation}
We call this choice the \emph{unit gauge} and impose it throughout. It fixes the associator when one of the three objects being reassociated is the vacuum, whereas the unitary gauge concerns preservation of inner products. The condition does not extend to $F_1^{abc}$: three non-vacuum objects can have total charge $1$ and still admit a non-trivial reassociation. Nor does it make entries with an intermediate vacuum channel trivial.

Entries with intermediate vacuum channels obey quantum-dimension constraints, which we call ``unit-row/column'' constraints. For example, consider the entry of $F_a^{a\bar a a}$ for which the intermediate channel is the vacuum in both fusion trees. In a unitary gauge its magnitude is fixed by the quantum dimension:
\begin{align}
    |[F_a^{a\bar a a}]_{(1,\circ,\circ),(1,\circ,\circ)}| = \frac{1}{d_a}
\end{align}
The quantum dimensions are known from the modular data, so this supplies a constraint before any consistency equations are solved. For self-dual $a$, the second Frobenius--Schur indicator also fixes the sign in the unit gauge:\footnote{For non-self-dual simples there exists a compatible gauge where $[F^{a\bar a a}_a]_{(1,\circ,\circ),(1,\circ,\circ)} = \frac1{d_a}$, but this gauge is not in general compatible with the later fixed-$R$ gauge choices.}
\begin{equation}
    [F^{a\bar a a}_a]_{(1,\circ,\circ),(1,\circ,\circ)} = \frac{\nu_2(a)}{d_a}
\end{equation}
where $\nu_2(a)=\pm1$~\cite{ngHigherFrobeniusSchurIndicators2007}. Applied to our self-dual generator $X$, this fixes a known entry of the initial matrix $F_X^{XXX}$ used in \cref{sec:floor}.

A braided category further possesses natural braiding isomorphisms $a\otimes b\cong b\otimes a$, which induce linear isomorphisms
\begin{align}
    R^{ab}_c:V_c^{ba}&\longrightarrow V_c^{ab}
    \label{eq:r-map}
\end{align}
given in components as
\begin{align}
    \vcenter{\hbox{\BraidedSplittingTree}}
    &=\sum_{\beta}
      [R_c^{ab}]_{\alpha\beta}
      \vcenter{\hbox{\ReorderedSplittingTree}}
    \label{eq:r-move}
\end{align}
The entries of these matrices are the $R$-symbols. They use the same junction bases as the $F$-symbols and obey $R^{ab}_c (R^{ab}_c)^\dagger = I$ in the unitary gauge described above.

The hexagon equations require that braiding an anyon past a composite object agree with braiding it past the two constituents, with the necessary reassociations. The two diagrams express this compatibility for the braiding and its inverse:
\begin{subequations}\label{eq:hexagon}
\begin{gather}
    \vcenter{\hbox{\HexagonCoherenceDiagram}}
    \label{eq:hexagon-plus}
    \displaybreak[0]\\[15pt]
    \vcenter{\hbox{\InverseHexagonCoherenceDiagram}}
    \label{eq:hexagon-minus}
\end{gather}
\end{subequations}
In components, these read
\begin{subequations}\label{eq:hexagon-components}
\begin{equation}
    \sum_{\lambda,\kappa}[R_e^{ba}]_{\beta\lambda}[F_d^{bac}]_{(e,\alpha,\lambda),(g,\mu,\kappa)}
[R_g^{ca}]_{\kappa\nu} = \sum_{f,\sigma,\tau,\eta} [F_d^{abc}]_{(e,\alpha,\beta),(f,\sigma,\tau)}[R_d^{fa}]_{\sigma\eta}[F_d^{bca}]_{(f,\eta,\tau),(g,\mu,\nu)} \label{eq:hexagon-components-plus}\end{equation}\begin{equation}
\sum_{\lambda,\kappa}[(R_e^{ab})^{-1}]_{\beta\lambda}[F_d^{bac}]_{(e,\alpha,\lambda),(g,\mu,\kappa)}
[(R_g^{ac})^{-1}]_{\kappa\nu}= \sum_{f,\sigma,\tau,\eta} [F_d^{abc}]_{(e,\alpha,\beta),(f,\sigma,\tau)}[(R_d^{af})^{-1}]_{\sigma\eta}[F_d^{bca}]_{(f,\eta,\tau),(g,\mu,\nu)} \label{eq:hexagon-components-minus}
\end{equation}
\end{subequations}
We will refer to the two hexagon systems with outgoing $a,b,c$ and incoming $d$ as $H^\pm(a,b,c;d)$.

A ribbon category also has a twist, describing a full rotation of an anyon. It is a natural isomorphism with components
\begin{equation}
    \theta_a : a \rightarrow a
\end{equation}
Abusing notation, we write $\theta_a = \theta_a\,\id_a$, where the scalar $\theta_a$ is a root of unity. Given a full set of $F$- and $R$-symbols, a choice of twists is equivalent to a choice of spherical structure (a pivotal structure for which left and right categorical traces agree)~\cite{etingofTensorCategories2015}. In the case of unitary modular categories, there is a canonical unitary ribbon structure~\cite{Galindo:2014}, given by
\begin{equation}
\theta_a = \frac{1}{d_a}\sum_{c} d_c\,\operatorname{Tr}(R^{aa}_c)
\end{equation}
Thus, in the unitary case the twists are fixed by the braiding and quantum dimensions.

The modular data are $(S,T)$, where $T_{ab}=\delta_{ab}\theta_a$ and $S_{ab}$ is the Hopf-link invariant labeled by $a,b$, divided by the total quantum dimension $D=(\sum_a d_a^2)^{1/2}$. We use the normalization $S_{1a}=d_a/D$, so $S_{11}=1/D$; $S$ is invertible in a modular category and unitary in the unitary case.

Two successive exchanges of $a$ and $b$ return them to their original order. In fusion channel $c$, this double exchange equals the twist of $c$ relative to the individual twists of $a$ and $b$. This is the balancing relation:
\begin{equation}
    R^{ba}_c R^{ab}_c
    =\frac{\theta_c}{\theta_a\theta_b}\,I
\end{equation}
To recover the individual exchange matrices, we use the gauge construction of Ng, Rowell, and Wen~\cite{ngRecoveringRsymbolsModular2024}. For distinct anyon labels, balancing and a choice of junction bases suffice. For identical labels, we must additionally determine the multiplicities of two allowed eigenvalues. Under changes of basis, the $R$-symbols transform as
\begin{equation}
    R^{ab}_c \mapsto U^{ba}_c R^{ab}_c (U^{ab}_c)^{-1}
\end{equation}
where $U^{ab}_c \in \mathrm{GL}(N^{ab}_c,\mathbb C)$ in general and $U^{ab}_c \in \mathrm{U}(N^{ab}_c)$ if unitarity is to be preserved.

First consider distinct labels. Fix an arbitrary total order on the simple objects with $1$ first. The bases can be chosen so that $R^{ab}_c = I$ for $a<b$, after which balancing fixes the reverse exchange to $R^{ba}_c = \frac{\theta_c}{\theta_a\theta_b}\,I$. Preserving this choice requires simultaneous basis changes, $U^{ba}_c = U^{ab}_c$ for $a\neq b$.

For identical labels, we choose a basis that diagonalizes $R^{aa}_c$. Balancing restricts its entries to $\pm \frac{\sqrt{\theta_c}}{\theta_a}$ for a fixed choice of $\sqrt{\theta_c}$. The two eigenvalue multiplicities sum to $N^{aa}_c$, and their difference is determined by $\operatorname{Tr} R^{aa}_c$. Thus, knowing this trace fixes the matrix up to the ordering of its entries.

The required traces are supplied by generalized Frobenius--Schur indicators~\cite{ngRecoveringRsymbolsModular2024,ngCongruenceSubgroupsGeneralized2010}:\footnote{Here $\iota:\mathcal C\to\mathcal Z(\mathcal C)$ sends $c$ to the same object equipped with the inverse-braiding half-braiding: its interchange $c\otimes a\to a\otimes c$ is the inverse of the braiding $a\otimes c\to c\otimes a$, following~\cite{ngRecoveringRsymbolsModular2024}. More generally, these indicators determine traces of powers of the braid rotation $C_n=\rho(\sigma_{n-1})\cdots\rho(\sigma_1)$ on $\Hom(c,a^{\otimes n})$, where $\rho$ is the braid group representation reviewed in \cref{sec:braid-representations}~\cite{Barter_2018}.}
\begin{equation}
    \nu_{2,1}^{\iota(c)}(a) = \theta_a \operatorname{Tr}(R^{aa}_c)
\end{equation}
These indicators can themselves be evaluated from the modular data~\cite{ngCongruenceSubgroupsGeneralized2010}:
\begin{equation}
    \nu_{m,1}^{\iota(c)}(a) = \frac{1}{D}\sum_{i,j} d_i \overline{S_{cj}} N^{ij}_a \left(\frac{\theta_i}{\theta_j}\right)^m
\end{equation}
Taking $m=2$ gives the trace needed above. Denoting the positive- and negative-sign eigenvalue multiplicities by $m_+$ and $m_-$, respectively, we obtain
\begin{equation}
    R^{aa}_c = \frac{\sqrt{\theta_c}}{\theta_a} \begin{pmatrix}
        I_{m_+} & 0\\0 & -I_{m_-}
    \end{pmatrix}, ~~~~m_\pm = \frac12\left(N^{aa}_c \pm \frac{1}{\sqrt{\theta_c} D} \sum_{i,j} d_i \overline{S_{cj}} N^{ij}_a \left(\frac{\theta_i}{\theta_j}\right)^2\right)
\end{equation}
Together, these choices fix all $R$-symbols in a unitary gauge using only the modular data.

With $R$ fixed, the hexagons become quadratic constraints on $F$ alone, involving fewer independent $F$-symbols than the cubic pentagon equations. They therefore provide tractable subsystems to solve first. Unlike the full pentagon system, however, the hexagons need not be rigid: they can leave continuous families of solutions even after gauge freedom is removed.

The unit gauge and the fixed $R$-matrices still leave a residual group of unitary basis changes:
\begin{equation}
    \mathcal G = \prod_{\substack{1<a<b\\c}} \mathrm{U}(N^{ab}_c) \times \prod_{\substack{a>1\\c}} \left(\mathrm{U}(m_+(a,c)) \times \mathrm{U}(m_-(a,c))\right) \label{eq:residual-gauge-group}
\end{equation}
The first product describes simultaneous changes of basis for distinct labels, preserving $U^{ba}_c = U^{ab}_c$. The second describes independent unitary rotations within the two eigenspaces of $R^{aa}_c$. Junctions with a vacuum outgoing leg are excluded because their bases have already been fixed by the unit gauge. Consequently, a continuous family found by a partial solve may represent either this residual basis freedom or genuinely undetermined data.

\subsection{Braid Group Representations}
\label{sec:braid-representations}

With the $R$-symbols fixed by the modular data, we seek constraints that determine manageable subsets of the unknown $F$-symbols. To obtain such constraints, we consider exchanges of several identical anyons: in a chosen fusion-tree basis, their exchange matrices combine the known $R$-symbols with the unknown $F$-symbols needed to reassociate the tree. These matrices must form a representation of the braid group, whose relations therefore constrain the $F$-symbols that enter them. Increasing the number of strands introduces further $F$-symbols while retaining the previously determined braid actions, providing the successive stages of our algorithm.

The braid group $B_n$ on $n$ strands is generated by $\sigma_1,\ldots,\sigma_{n-1}$, where $\sigma_i$ exchanges the $i$-th and $(i+1)$-th strands, and has the presentation
\begin{align}
    B_n=
    \left\langle
    \sigma_1,\ldots,\sigma_{n-1}
    \;\middle|\;
    \begin{aligned}
        \sigma_i\sigma_{i+1}\sigma_i
        &=\sigma_{i+1}\sigma_i\sigma_{i+1}
        && (1\leq i\leq n-2),\\
        \sigma_i\sigma_j&=\sigma_j\sigma_i
        && (|i-j|>1)
    \end{aligned}
    \right\rangle
    \label{eq:braid-relations}
\end{align}
The first relation is the braid form of the Yang--Baxter equation; the second expresses commutativity of exchanges on disjoint pairs of strands.

For simple objects $X$ and $t$, the space $\Hom(t,X^{\otimes n})$ consists of junctions splitting total charge $t$ into $n$ copies of $X$. Braiding these identical outgoing anyons acts on this space~\cite{etingofTensorCategories2015,Nayak:2008zza,rowellMathematicsTopologicalQuantum2018}. In a unitary gauge this gives a unitary representation
\begin{equation}
    \rho_{n,t}:B_n\longrightarrow \mathrm{U}(\Hom(t,X^{\otimes n}))
\end{equation}
We use the left-associated fusion-tree basis. The label $a_k$ is the cumulative charge of the first $k$ copies of $X$, and $\mu_k$ labels the junction at which that charge is resolved into $a_{k-1}$ and the next $X$. Thus the basis states are
\begin{equation}
\ket{\mathbf a,\boldsymbol\mu} =\ket{a_2,\ldots,a_{n-1};\mu_2,\ldots,\mu_n},~~~~~1\leq\mu_k \leq N^{a_{k-1}X}_{a_k},~~~~~a_1=X,~a_n=t
\end{equation}
In this basis the first two strands already meet at a single junction. The generator $\sigma_1$ therefore acts on $(\dots(X\otimes X)_{a_2}\dots)\rightarrow t$ through $R_{a_2}^{XX}$, leaving all other junction labels unchanged:
\begin{equation}
    \bra{\mathbf a',\boldsymbol\mu'} \rho_{n,t}(\sigma_1) \ket{\mathbf a,\boldsymbol\mu} = [R_{a_2}^{XX}]_{\mu_2'\mu_2}
\left(\prod_{k=2}^{n-1}\delta_{a_k'a_k}\right)
\left(\prod_{k=3}^{n}\delta_{\mu_k'\mu_k}\right)
\end{equation}
For $i>1$, the two strands to be exchanged do not meet at a single junction in the chosen tree. We first reassociate the subtree $((a_{i-1}\otimes X)_{a_i}\otimes X)_{a_{i+1}}$ to $(a_{i-1}\otimes(X\otimes X)_b)_{a_{i+1}}$, braid the two $X$ strands, and then reassociate back. On each block with all labels other than $a_i,\mu_i,\mu_{i+1}$ fixed, this gives
\begin{equation}
    \rho_{n,t}(\sigma_i) =
F^{a_{i-1}XX}_{a_{i+1}}
\left[
\bigoplus_b
\left(I_{N^{a_{i-1}b}_{a_{i+1}}}\otimes R_b^{XX}\right)
\right]
(F^{a_{i-1}XX}_{a_{i+1}})^{-1}
\end{equation}
The direct sum runs over the possible charges $b$ of the exchanged pair. The identity factor leaves the junction joining $a_{i-1}$ and $b$ unchanged. In components,
\begin{multline}\label{eq:braid-reps}
    \bra{\mathbf a',\boldsymbol\mu'} \rho_{n,t}(\sigma_i) \ket{\mathbf a,\boldsymbol\mu}
=\\\sum_{b,\kappa,\kappa',\lambda}
\left[F^{a_{i-1}XX}_{a_{i+1}}\right]_{
(a_i',\mu_{i+1}',\mu_i'),(b,\lambda,\kappa')}
\times
[R_b^{XX}]_{\kappa'\kappa}
\times
[(F^{a_{i-1}XX}_{a_{i+1}})^{-1}]_{
(b,\lambda,\kappa),(a_i,\mu_{i+1},\mu_i)} \\ \times\left(
\prod_{\substack{2\le k\le n-1\\k\ne i}}
\delta_{a_k'a_k}
\right)
\left(
\prod_{\substack{2\le k\le n\\k\ne i,i+1}}
\delta_{\mu_k'\mu_k}
\right)
\end{multline}
The Kronecker deltas keep all junction labels outside the reassociated subtree fixed. Substituting these matrices into the braid relation $\sigma_i\sigma_{i+1}\sigma_i = \sigma_{i+1}\sigma_i\sigma_{i+1}$ gives non-linear constraints involving only the $F$-symbols $F^{aXX}_b$, with the $R$-symbols already known.

Although the hexagon equations~\eqref{eq:hexagon}, together with naturality, imply the braid relations, the latter constrain smaller subsystems of $F$-symbols. This makes them useful starting points before imposing the stronger hexagon conditions.

The braid groups also organize the successive stages of the algorithm. Adding strands on which a braid acts trivially gives, for $m>n$, the canonical embedding
\begin{equation}
    \iota_{n,m}:B_n\hookrightarrow B_m,
~~~~~
\sigma_i\mapsto\sigma_i,
~~~~ 1\le i\le n-1
\end{equation}
To see how this works on junction spaces, consider $n+1$ strands with total charge $t$. The first $n$ strands have an intermediate total charge $a$, which combines with the last $X$ through one of $N^{aX}_t$ junctions. Braiding only the first $n$ strands leaves this final junction untouched. The left-associated basis therefore gives $\rho_{n+1,t}\circ\iota_{n,n+1}=\bigoplus_a\rho_{n,a}\otimes I_{N^{aX}_t}$. Thus the already determined representations on $n$ strands fix the first $n-1$ generators on $n+1$ strands, and only the image of the new generator $\sigma_n$ remains to be found.

\subsection{Exotic Rank-11 Anisotropic Data}
\label{sec:rank-eleven}

Our running example is the completely anisotropic, pseudo-unitary (all $d_a>0$) rank-11 modular data \href{https://anyonatlas.org/data/B11-00119}{\texttt{B11-00119}} in \href{https://anyonatlas.org}{\texttt{AnyonAtlas.org}}~\cite{anyonAtlas}. No categorification was known when these data were first identified in~\cite{ngClassificationModularData2025}. Their possible significance for unitary Witt classes beyond Chern--Simons theory was discussed in~\cite{Berean-Dutcher:2026fck} and in the introduction; here we describe the features relevant to the computation.

We label the simple objects of a modular category realizing \texttt{B11-00119} as $V_0=1,V_1,\ldots,V_{10}$, in the order of the modular data displayed in~\cite[Sec.~5]{Berean-Dutcher:2026fck}. Each of the ten non-vacuum simples is self-dual and generates the category, so there are several possible starting points. We take $X=V_1$, whose first tensor powers are
\begin{align}
    X^{\otimes2}&=1+X+V_2+V_4+V_5+V_6+V_9,\\
    X^{\otimes3}&=1+7X+3V_2+V_3+5V_4+5V_5+6V_6+3V_7+5V_8+7V_9+6V_{10}
\label{eq:rank-eleven-generator}
\end{align}
Thus $X^{\otimes 3}$ contains every simple object, while $X\otimes X$ is multiplicity-free. These features give a relatively simple starting point before the ladder reaches junctions with higher multiplicities. More generally, when several generators are available, we choose one for which the successive subsystems grow in difficulty as gradually as possible. We write $a\in X^{\otimes n}$ when $a$ occurs as a simple summand; unions and differences of tensor powers below refer to these sets of summands, ignoring multiplicities.

The full fusion rules are not multiplicity-free: already $N^{V_9X}_{V_9}=2$, and the largest fusion coefficient is $4$ (e.g.\ $N^{V_9V_9}_{V_9}$). In total there are 10,886 non-trivial $F$-matrices, by which we mean nonempty matrices $F^{abc}_d$ with $a,b,c\neq1$, containing 3,855,922 $F$-symbols. This example therefore tests the method well beyond the multiplicity-free setting, despite the simplicity of the first tensor square.

\section{\texorpdfstring{$F$}{F}-Symbol Algorithm}
\label{sec:alg}

We now describe an algorithm for finding complete sets of $F$-symbols compatible with given modular data, and for reconstructing and certifying their exact forms. The algorithm climbs a ladder of braid group representations on successively more strands of a self-dual simple generator $X$, using the embeddings described in \cref{sec:braid-representations}. Each rung adds one strand and introduces a restricted family of new $F$-symbols to determine.

At each rung, we use the braid, hexagon, and pentagon equations to constrain the new $F$-symbols, choosing subsystems that introduce as few additional unknowns as possible. A partial solution need not yet be fixed up to gauge: the $F$-symbols remain free to vary as further constraints are added, until their non-gauge moduli have been removed. The aim is to determine enough $F$-symbols to obtain the rest by propagation, confining the non-linear searches to a small part of the full problem. This use of smaller, underdetermined constraint systems to make larger solves tractable parallels the strategy in~\cite{Berean-Dutcher:2026fck}, where Frobenius--Schur indicator constraints are imposed before solving the Verlinde equations for modular data.

After stating our assumptions in \cref{sec:assumptions}, we describe the ladder and subsequent propagation in \cref{sec:braid-group-ladder}, using the exotic rank-11 category of \cref{sec:rank-eleven} as a guiding example. We explain how to reconstruct exact $F$-symbols from a numerical solution in \cref{sec:reconstruction} and how to certify every pentagon and hexagon equation in \cref{sec:certification}.

\subsection{Assumptions}
\label{sec:assumptions}

The key assumption of our algorithm is:

\par\medskip
\noindent\textbf{Assumption 1} (Self-dual generator): There exists a self-dual simple object $X$ that generates $\C$, i.e.\ every simple object is a summand of $X^{\otimes n}$ for some $n$.\par\medskip
This is the structural input for the ladder: the junction spaces $\Hom(t,X^{\otimes n})$ can reach every simple object $t$ for sufficiently large $n$. Self-duality supplies additional constraints, including the vacuum-channel entry of $F_X^{XXX}$ fixed by the Frobenius--Schur indicator at the floor. Of the 3,220 pseudo-unitary modular data through rank 22 in~\cite{Berean-Dutcher:2026fck}, we find that 1,669 have such a generator.

Our numerical implementation and several algebraic steps also assume:
\par\medskip
\noindent \textbf{Assumption 2} (Unitarity): The $F$- and $R$-symbols admit a unitary gauge.\par\medskip
This allows us to parameterize the unknown $F$-symbols by unitary matrices. Unitarity also allows certain pentagon systems with two unknown $F$-matrices to be rewritten as linear constraints, as described in \cref{sec:ceiling}. It is also physically motivated, since unitary modular categories describe the anyons of gapped bosonic topological phases~\cite{kitaevAnyonsExactlySolved2006,Nayak:2008zza}. Unitarity requires the modular data to be pseudo-unitary, but positivity of the quantum dimensions alone is not known to establish the existence of a unitary categorification.

Both assumptions can likely be relaxed, at the cost of a less efficient non-linear solve. For example, the ladder could be built on a pair of simple generators of the fusion ring, with more $F$-matrices needed before propagation. Similarly, the search could be performed over $\mathrm{GL}(n,\mathbb C)$ instead of $\mathrm{U}(n)$. The pentagon steps that rely on unitarity for linearity would then, in some cases, need to solve for inverse $F$-matrices instead, which is by no means intractable. The effect of these changes on the computational cost remains to be assessed.

Finally, to reconstruct the exact form of the $F$-symbols, we assume:
\par\medskip
\noindent\textbf{Assumption 3} (Cyclotomic field): There exists a gauge compatible with the fixed $R$-symbols in which the $F$-symbols lie in $\mathbb Q(\zeta_{2\ord(T)})$, where $\zeta_n = e^{2\pi i/n}$ and $\ord(T)$ is the least positive integer $q$ such that $T^q=I$.\par\medskip
The numerical search itself does not require a choice of number field. By Ocneanu rigidity, the $F$-symbols always lie in some number field in a suitable gauge~\cite{etingofFusionCategories2017}, but this does not guarantee a gauge over a prescribed field. The additional assumption above is used to recover exact values from the numerical solution. The reconstruction gauge need not be unitary, as explained in \cref{sec:reconstruction}. Although the chosen cyclotomic field contains the fixed $R$-symbols, it need not contain the $F$-symbols in any compatible gauge. Indeed, Gannon, Schopieray and Yadav~\cite{Gannon:2026enh} prove that some modular categories admit no gauge in which both $F$- and $R$-symbols lie in a cyclotomic field; the Drinfeld center of a $\zee{}/5\zee{}$ Haagerup--Izumi category is one example. The cyclotomic choice is not essential to the strategy: reconstruction could be adapted to another suitable number field $K$ containing the fixed $R$-symbols and the $F$-symbols in a compatible gauge. The existence of such a gauge for a proposed $K$, however, is generally not known in advance.

\subsection{Braid Group Ladder}
\label{sec:braid-group-ladder}

The main engine for our algorithm is described here and illustrated in \cref{fig:braid-group-ladder}. We first solve the highly constrained matrix $F^{XXX}_X$, which forms the floor on which our ladder stands (\cref{sec:floor}). Each subsequent rung adds one strand of $X$ and solves for the $F$-symbols that enter the braid group representations on $\Hom(t,X^{\otimes n})$; the final step, the ceiling (\cref{sec:ceiling}), propagates the solution to all remaining $F$-symbols. In our guiding rank-11 example, we need only go to $n=5$ before enough $F$-symbols have been determined to allow propagation to a complete solution. The number of rungs needed varies from case to case, and further constraints may be required (see \cref{sec:disc}).

\begin{figure}[htbp]
    \centering
    \includegraphics[width=\linewidth]{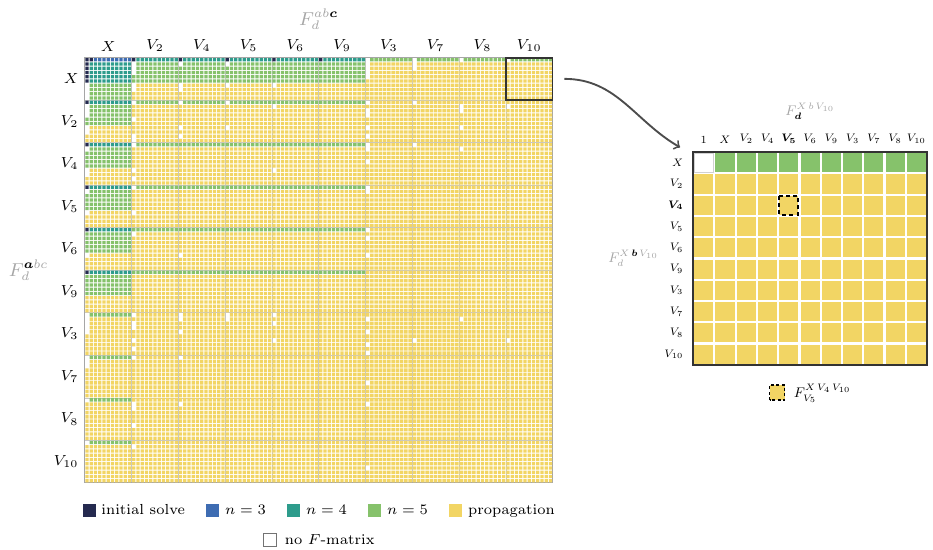}
    \caption{Stages of the $F$-symbol calculation for the rank-11 category. Each colored cell represents one non-trivial $F$-matrix $F^{abc}_d$, with its color indicating the algorithm stage. Rows are labeled by $(a,b)$ and columns by $(c,d)$. Objects are ordered by their first occurrence in powers of $X$ (see the fusion rules in \cref{eq:rank-eleven-generator}), with $d=1$ first in each group. Blank cells indicate that there is no such $F$-matrix, i.e.\ that $\Hom(d,a\otimes b\otimes c)=0$. The initial solve and the $n=3,4,5$ rungs supply 1,106 matrices, from which the remaining 9,780 are obtained by propagation. These 1,106 matrices contain only 3.4\% of all scalar $F$-symbols.}
    \label{fig:f-symbol-stages}
\end{figure}

\begingroup\interlinepenalty=10000
\Cref{fig:f-symbol-stages} shows how these stages fill out the $F$-matrices of the rank-11 example: the initial solve and successive rungs determine matrices containing only 3.4\% of the scalar $F$-symbols, from which propagation recovers the remainder.\par
\endgroup

\subsubsection{Floor}
\label{sec:floor}

Climbing the ladder requires an entry point. At later rungs, many of the $F$-symbols in the constraints are already known; at the floor none are, so the entry point should be a small, highly constrained subsystem. We take the single matrix $F_X^{XXX}$, which, although relatively large, can be tightly constrained without introducing many additional $F$-symbols. We do so with the system
\begin{equation}
    H^\pm(X,X,X;X) \cup P(X,X,X,X;1)\cup H_\text{vac} \cup V \label{eq:floor-system}
\end{equation}
where $V$ is the dimension constraint
\begin{equation}
    [F_X^{XXX}]_{(1,\circ,\circ),(1,\circ,\circ)} = \frac{\nu_2(X)}{d_X}
\end{equation}
and $H_\text{vac}$ is the set of vacuum hexagons
\begin{equation}
    H^\pm(X,X,X;1) \cup H^\pm(e,X,X;1) \cup H^\pm(X,e,X;1) \cup H^\pm(X,X,e;1)
\end{equation}
for $e\in X^{\otimes 2}$. The vacuum pentagon $P(X,X,X,X;1)$ also introduces $F^{XXX}_1$ as well as $F^{eXX}_1$, $F^{XeX}_1$, and $F^{XXe}_1$ for the simples $e\in X^{\otimes 2} \setminus \{1,X\}$ (the matrices with $e=1$ are trivial in the unit gauge of \cref{eq:unit-gauge}, e.g.\ $F^{XX1}_1 = 1$). We therefore include the vacuum hexagons for these matrices as well. In the rank-11 case, $F_X^{XXX}$ is a $7\times 7$ matrix, and the 16 additional $F$-symbols are each a single complex variable, for a total of 65 complex variables at this stage in the solver, or 65 real parameters after imposing unitarity.

We solve this subsystem numerically in a unitary gauge. The multistart non-linear solvers are initialized with Haar-distributed random unitary $F$-matrices, which we term cold starts. At a given solution, we count the moduli (non-gauge null directions) as the Jacobian nullity minus the dimension of the local gauge orbit. Subtracting the dimension of the residual gauge group itself would generally overcount the gauge directions, since part of the group may act trivially on the solution. Although this count is not reliable at singular points, it suffices for our purposes. In later non-linear solves, $F$-symbols not yet fixed up to gauge are warm-started: they are initialized at the previous solution but allowed to vary. Those with no remaining moduli are frozen, reducing the number of variables.

For the rank-11 category, the floor system fixes the solution up to one modulus, with 11 local gauge directions. Additional constraints are therefore needed to fully determine these $F$-symbols. In general, we could simply pass the underdetermined solution to the next rung, adding constraints until no moduli remain. When possible, however, we add conditions that reduce the number of moduli. For the rank-11 category, it suffices to re-solve with the additional Gram identity~\cite{bondersonNonAbelianAnyons2007}
\begin{equation}\label{eq:unit-row2}
    M_f M_f^\dagger = \frac{d_f}{d_X^2} I,\qquad (M_f)_{\mu\nu} \coloneq [F_X^{XXX}]_{(1,\circ,\circ),(f,\mu,\nu)}
\end{equation}
for each $f\in X^{\otimes 2}$, which eliminates the remaining modulus.

\subsubsection{\texorpdfstring{$n=3$}{n=3} Rung}

At the first rung of the ladder, we solve for $\rho_{3,t}(\sigma_2)$, the image of the generator $\sigma_2$ of $B_3$ acting on $\Hom(t,X^{\otimes 3})$; the image of $\sigma_1$ is already fixed by the $R$-symbols in our gauge. Since the hexagons below imply the braid relations of $B_3$ and involve no further $F$-symbols, we do not solve the braid relations themselves, but instead use them to identify the relevant $F$-matrices: $F_t^{XXX}$ for $t\in X^{\otimes 3}$. In the rank-11 case, every simple object occurs in $X^{\otimes 3}$, so all 11 blocks are considered, with sizes ranging from 1 to 7; of these, $F^{XXX}_1$ and $F_X^{XXX}$ are already solved. We add to the earlier system the hexagons
\begin{equation}
    \bigcup_{t\in X^{\otimes 3}} H^\pm(X,X,X;t)
\end{equation}
which introduce no additional $F$-symbols and constrain each $F_t^{XXX}$ matrix independently. No pentagons coupling these blocks are added at this stage, as they would generally introduce further unknown $F$-symbols.

The previously solved matrices $F^{XXX}_1$ and $F_X^{XXX}$ remain free to vary while solving the enlarged system. For the rank-11 category, the solution has 50 null directions, 47 of which are gauge, leaving at most 3 moduli.

The simple representations of $B_3$ of dimension at most 5 were classified in~\cite{tuba1999representationsbraidgroupb3}. This classification could be used to identify the small representations exactly. However, the $n=3$ hexagon system is not the most challenging phase of the algorithm, and larger braid representations often appear there, so we have not used this classification.

\subsubsection{\texorpdfstring{$n=4$}{n=4} Rung}

Having now determined the action of $\sigma_1$ and $\sigma_2$ on the spaces $\Hom(a,X^{\otimes 3})$ and embedded it in $\Hom(t,X^{\otimes 4})$, we are left with only one undetermined generator of $B_4$, namely $\sigma_3$. Its representation is built from the matrices $F_t^{eXX}$ for $e\in X^{\otimes 2}$ and $t\in X^{\otimes 4}$. Since our guiding principle is to solve the smallest available subsystems first, this time we impose the braid relations before the hexagons, which would introduce the further unknowns $F_t^{XeX}$ and $F_t^{XXe}$. From the braid group representations in \cref{eq:braid-reps}, we solve the braid relation $\sigma_2\sigma_3\sigma_2 = \sigma_3\sigma_2\sigma_3$ for the undetermined $\rho_{4,t}(\sigma_3)$, for all $t\in X^{\otimes 4}$.

Since the braid generators involve $R$-symbols conjugated by $F$, when $R$ has repeated eigenvalues, there will be additional null directions in the $F$-symbol solutions that need not be gauge. To limit these, we impose two further constraints. First, we identify the block $e=X$ with the matrix $F_t^{XXX}$ already found on the $n=3$ rung. Second, if $t\in X^{\otimes 2}$, we can impose a Gram identity generalizing \cref{eq:unit-row2}~\cite{bondersonNonAbelianAnyons2007}. Namely, for
\begin{equation}
    (M_f)_{\mu\nu} \coloneq [F_t^{tXX}]_{(f,\mu,\nu),(1,\circ,\circ)}
\end{equation}
we impose
\begin{equation}
    M_f M_f^\dagger = \frac{d_f}{d_td_X} I
\end{equation}
Once these constraints remove the moduli of the previous $F_t^{XXX}$ matrices, those matrices can be held fixed.

With $F_t^{eXX}$ determined, the remaining $F_t^{XeX}$ and $F_t^{XXe}$ are relatively straightforward to solve. We can do this in two ways, using either the pentagon $P(X,X,X,X;t)$ or the hexagons $H^\pm(e,X,X;t)$. Both are eventually imposed, but it is preferable to solve one before imposing the other. We solve the pentagons first, as they are more constraining and not substantially more challenging to solve. Note that this is the first pentagon used since the floor; the algorithm is tractable because it never attempts to solve the full pentagon system directly. With all but two of the $F$-matrices in this pentagon system determined, each on opposite sides, the resulting constraints are jointly linear in the unknowns, of the form
\begin{equation}
    W = LVR
\end{equation}
with $L,R$ known. The block structures of the $F$-matrices alone constrain the majority of this linear system.

At this point, all $F_t^{eXX}$, $F_t^{XeX}$, and $F_t^{XXe}$ matrices have been solved for $e\in X^{\otimes 2}\cup\{X\}$ and $t\in X^{\otimes 4}$. For the rank-11 category, this is almost all we need. 

\subsubsection{\texorpdfstring{$n\geq 5$}{n>=5} Rungs}

The process now repeats. Once $B_4$ is embedded in $B_5$, only the generator $\sigma_4$ of $B_5$ remains undetermined. Its representation is built from the matrices $F_t^{aXX}$ for $a\in X^{\otimes 3}$, of which only $F_t^{bXX}$ with $b\in X^{\otimes 3} \setminus X^{\otimes 2}$ are still unknown. We solve for these, as before, using the braid relations. To avoid additional moduli due to repeated $R$-eigenvalues, we impose the zig-zag constraint for the self-dual $X$ (cf.~\cite[App.~E]{kitaevAnyonsExactlySolved2006}):
\begin{equation}
    \sum_{\lambda} [F_b^{bXX}]_{(a,\mu,\lambda),(1,\circ,\circ)}[(F_a^{aXX})^\dagger]_{(1,\circ,\circ),(b,\lambda,\mu')} = \frac{\nu_2(X)}{d_X}\delta_{\mu,\mu'}
\end{equation}
for $b\in X^{\otimes 3}\setminus X^{\otimes 2}$ and $a\in X^{\otimes 3}$.

Having completed the braid relations, we again pass to the pentagons. These introduce more new $F$-matrices than at the $n=4$ rung, and so are solved in a particular order. First, the $P(e,X,X,X;t)$ pentagons are solved for each $e\in X^{\otimes 2}$. In addition to the already fixed $F$-symbols, these involve $F^{eXj}_t$ and $F^{ejX}_t$ for $j\in X^{\otimes 2}$. We then solve $P(X,e,X,X;t)$, which introduces $F_t^{Xej}$ and $F_t^{XbX}$ for $b\in X^{\otimes 3}\setminus X^{\otimes 2}$. Then, $P(X,X,e,X;t)$ introduces only $F^{XXb}_t$, since the $F^{jeX}_t$ it contains are among the $F^{ejX}_t$ already found from $P(e,X,X,X;t)$. Finally, $P(X,X,X,e;t)$ introduces no new $F$-symbols: the $F^{jXe}_t$ and $F^{Xje}_t$ it contains are already determined by the first two pentagons. As before, each of these pentagons has at most one unknown $F$-matrix on each side and so gives linear constraints.

At this point, enough $F$-symbols have been determined, with no remaining moduli, to complete the solution for the rank-11 category. The point at which this occurs varies category by category, with some requiring additional rungs, which can be implemented by the same procedure.

Were one to proceed higher in rank, additional $F$-symbol constraints such as those used previously might be needed to reduce the moduli introduced by solving hexagons or braid relations. For this purpose, the tetrahedral symmetries described in~\cite{fuchsTetrahedralSymmetry6jsymbols2021} may be of use. These are relations describing the action of $S_4$ on the $F$-symbols of fusion categories. Once an apparent normalization issue in Eq.~(4.21) of~\cite{fuchsTetrahedralSymmetry6jsymbols2021} is corrected, these relations take the form 
\begin{subequations}
\begin{align}
[(F_a^{bcd})^{-1}]_{(e,\mu,\nu),(f,\rho,\sigma)}
&= \frac{d_b d_e}{d_a}
\sum_{\mu',\sigma'}
[(F_e^{\bar b b e})^{-1}]_{(a,\mu',\mu),(1,\circ,\circ)}
\notag\\
&\qquad\qquad\qquad\qquad\times
[F_c^{\bar b b c}]_{(1,\circ,\circ),(f,\sigma',\sigma)}
[F_e^{\bar b f d}]_{(c,\nu,\sigma'),(a,\mu',\rho)}
\\[1em]
[(F_a^{bcd})^{-1}]_{(e,\mu,\nu),(f,\rho,\sigma)}
&= d_c \lambda_{\bar c}
\sum_{\nu',\sigma'}
[(F_d^{\bar c c d})^{-1}]_{(e,\nu',\nu),(1,\circ,\circ)}
\notag\\
&\qquad\qquad\qquad\qquad\times
[(F_b^{b c \bar c})^{-1}]_{(1,\circ,\circ),(f,\sigma',\sigma)}
[F_a^{f \bar c e}]_{(b,\mu,\sigma'),(d,\rho,\nu')}
\\[1em]
[(F_a^{bcd})^{-1}]_{(e,\mu,\nu),(f,\rho,\sigma)}
&= \frac{d_d d_f}{d_a}
\sum_{\nu',\rho'}
[F_c^{c d\bar d}]_{(e,\nu',\nu),(1,\circ,\circ)}
\notag\\
&\qquad\qquad\qquad\qquad\times
[(F_f^{f d \bar d})^{-1}]_{(1,\circ,\circ),(a,\rho',\rho)}
[F_f^{b e \bar d}]_{(a,\rho',\mu),(c,\sigma,\nu')}
\end{align}
\end{subequations}
where $\lambda_{\bar c} = d_c
[F_{\bar c}^{\bar c c\bar c}]_{(1,\circ,\circ),(1,\circ,\circ)} \in \Uone$ and $\lambda_c = \nu_2(c)$ if $c$ is self-dual. The first relation implements the transposition $(12) \in S_4$, the second $(23)$, and the third $(34)$. Stronger forms of these symmetries exist for subclasses of modular categories; see, for example,~\cite {levinStringnetCondensationPhysical2005,Turaev2010}. These forms, however, generally require gauge choices incompatible with diagonal $R$-symbols. 
We emphasize, though, that these tetrahedral relations were not needed for the rank-11 example; we include them because they might assist in more difficult cases.

\subsubsection{Ceiling}
\label{sec:ceiling}

All that remains is to propagate our solution to the rest of the $F$-symbols. This is done in waves. Each wave first solves every hexagon and pentagon that is linear in a single unknown $F$-matrix. Then, in unitary categories, any pentagon and hexagon with two unknown $F$-matrices can be rewritten as jointly linear systems constrained by the block forms of the $F$-matrices, as at the earlier rungs. The lack of remaining moduli is verified during propagation. Repeating this process in waves allows the rest of the $F$-symbols to be determined starting from only a relatively small seed. In the rank-11 case, 1,106 of 10,886 non-trivial $F$-matrices had been solved before propagation, corresponding to only 132,739 of 3,855,922 scalar $F$-symbols. For exact reconstruction, we select a smaller seed, as described in \cref{sec:reconstruction}.

\subsection{Reconstruction}
\label{sec:reconstruction}

With a complete numerical solution to the $F$-symbols, we would like to recover their exact forms. For this, we assume that a gauge exists in which they lie in a cyclotomic field, namely $\mathbb Q(\zeta_{n})$ for $n=2\ord(T)$. This is not always possible~\cite{Gannon:2026enh}, but in most of the instances tested, we are able to recover $F$-symbols in such a gauge. We do not assume that this gauge is unitary; in fact, we relax the unitary gauge used previously so that the exact forms can be determined in a more convenient gauge.

The exact reconstruction of modular data in~\cite[Sec.~3.4]{Berean-Dutcher:2026fck} exploits the cyclotomic integrality of the unnormalized $S$ matrix and its known Galois action; we know of no comparable structure for the $F$-symbols, and so reconstruct them more directly. Knowing the full solution, fixed up to gauge, we can select a smaller reconstruction seed than that used as the input to the propagation in \cref{sec:ceiling}. We are not aware of a universal minimal set of $F$-symbols for this, but in the rank-11 case we need only reconstruct 216 matrices, or 12,188 $F$-symbols (just over 0.3\% of the total). Specifically, these are the 176 $F_t^{eXX}$, $F_t^{XeX}$, and $F_t^{XXe}$ for $e\in X^{\otimes 2}$ and the 40 $F_t^{bXX}$ for $b\in X^{\otimes 3}\setminus X^{\otimes 2}$.

Having chosen a seed, we refine its solution to high precision (1,200 to 1,500 digits in the rank-11 case). Residual gauge freedom is then exhausted by sending as many $F$-submatrices in the seed as possible to the identity. We expand the gauge group to admit non-unitary gauges, replacing each factor $\mathrm{U}(k)$ of $\mathcal G$ by $\mathrm{GL}(k,\mathbb C)$, but maintain the fixed $R$-symbols and the unit-gauge constraints on $F$. Whenever the gauge freedom at a single vertex $V^{ab}_c$ is the only one acting on an invertible square $F$-submatrix, we use it to fix that submatrix to the identity. This gauge-fixing process proceeds incrementally until the remaining gauge freedom acts trivially on the seed.\footnote{This gauge-fixing procedure is comparable to that used previously to solve for the $F$-symbols of low-rank multiplicity-free fusion categories in~\cite{vercleyen2024lowrankmultiplicityfreefusioncategories}, although here it serves a different purpose.}

We then use integer-relation detection to recognize the non-identity $F$-symbols as elements of $K=\mathbb Q(\zeta_{n})$, which is $\mathbb Q(\zeta_{70})$ in the rank-11 case. For $\zeta = \zeta_{n}$, consider the maximal real subfield $K^+ = \mathbb Q(\zeta + \zeta^{-1})$, over which $K$ has degree 2. Every $z\in K$ can be written uniquely as 
\begin{equation}
    z= x+y(\zeta - \zeta^{-1}),\qquad x,y\in K^+
\end{equation}
Since $x$ and $y$ are real, they are obtained numerically as $x=\Re z$ and $y=\Im z/(2\sin(2\pi/n))$. Then, using the basis $b_k = \zeta^k + \zeta^{-k}$, $0\leq k<\varphi(n)/2$, of $K^+$, with $\varphi$ Euler's totient function, we search for small integers $p_k$ and $q_k$ such that
\begin{equation}
    \sum_k (q_k x + p_k) b_k = 0
\end{equation}
For $q = \sum_k q_k b_k \neq 0$, this gives a candidate
\begin{equation}
    x = - \frac{\sum_k p_k b_k}{q}
\end{equation}
and $y$ is recognized in the same way. This rational recognition procedure continues until exact forms of the full seed are determined that satisfy the necessary constraints exactly.

Once this seed is reconstructed exactly, we repeat the propagation stage described previously, this time with the exact forms. Although this final step is conceptually simple, it naively involves propagating millions of elements of $K$ with increasingly large denominators. To avoid this, we instead perform much cheaper computations in finite fields $\mathbb F_p$ for primes $p \equiv 1 \pmod{n}$. The procedure we use is standard~\cite{Wang:1982} and, although more powerful versions exist, suffices here. Let $\mathcal O_K=\mathbb Z[\zeta]$ be the ring of integers of $K$. Since $p\equiv 1\pmod{n}$, the principal ideal generated by $p$ splits completely as
\begin{equation}
    p\mathcal O_K = (p) = \prod_{a\in (\zee{}/n\zee{})^\times}\mathfrak p_{p,a}
\end{equation}
where $\omega_p\in\mathbb Z$ reduces to a primitive $n$-th root of unity in $\mathbb F_p$, and the prime ideals $\mathfrak p_{p,a} = (p, \zeta - \omega_p^a)$ are the kernels of the evaluation maps
\begin{equation}
    \sigma_{p,a} : \mathcal O_K \rightarrow \mathbb F_p, ~~~~~\zeta\mapsto \omega_p^a
\end{equation}
The full propagation is then repeated for each of these $\varphi(n)$ reductions, yielding $x \mod \mathfrak p_{p,a}$ for every $F$-symbol $x$ and every prime ideal above $p$. By the Chinese Remainder Theorem,
\begin{equation}
\mathcal O_K/p\mathcal O_K
\cong
\prod_{a\in(\mathbb Z/n\mathbb Z)^\times}
\mathcal O_K/\mathfrak p_{p,a}
\end{equation}
so we can recover $x \mod p\mathcal O_K$. For this reduction to be well-defined, we must reject bad primes. Define the denominator ideal
\begin{equation}
    \mathcal D_{\mathcal E}
    = \{a\in\mathcal O_K : a\mathcal E\subseteq\mathcal O_K\}
\end{equation}
for $\mathcal E\subset K$, the set of seed entries.
We first require $p\nmid N_{K/\mathbb Q}(\mathcal D_{\mathcal E})$,
where $N_{K/\mathbb Q}(\mathfrak a)=[\mathcal O_K:\mathfrak a]$ denotes the norm of an integral ideal, ensuring that the seed admits reduction modulo every
$\mathfrak p_{p,a}$. For each linear system
$Lx=b$, with characteristic-zero nullity $g=\dim_K\ker L$,
we additionally require
\begin{equation}
    \dim_{\mathbb F_p}\ker \sigma_{p,a}(L)=g.
\end{equation}
For square $L$ with $g=0$, this reduces to $\det\sigma_{p,a}(L)\neq 0$. Repeating the propagation for sufficiently many good primes $\{p_i\}_{i=1}^k$ and applying the Chinese Remainder Theorem again, we can reconstruct $x \mod M\mathcal O_K$ for $M = \prod p_i$. If
\begin{equation}
    x = \left(\sum_{i=0}^{\varphi(n)-1} a_i\zeta^i\right)/b
\end{equation}
with integers $a_i$ and $b$ satisfying $|a_i| \leq A$ and $0<b \leq B$, then rational reconstruction recovers $x$ uniquely once $M>2AB$. In the rank-11 case, 100 primes were used, so the full propagation is performed 2,400 times. In the end, the $F$-symbols are determined exactly in a fixed gauge, comprising 92,542,128 rational coefficients in total. Once this exact form is found, an exact unitary gauge can be recovered by finding positive-definite Gram matrices $G^{ab}_c$ for Hermitian inner products on the junction spaces $V^{ab}_c$ that make all $F$- and $R$-maps isometries, and then passing to orthonormal bases using their positive square roots.

\subsection{Certification}
\label{sec:certification}

We now verify that the reconstructed $F$-symbols, together with the fixed $R$-symbols, satisfy every pentagon and hexagon equation exactly. Evaluating every equation directly in $K$ would again require enormous cyclotomic computations. Instead, we again work over sufficiently many finite fields $\mathbb F_p$. Schematically, we use the fact that if a residual $E$ of a pentagon or hexagon equation is divisible by some $M$, and we have a bound $|E| \leq H$ for $H<M$, then, necessarily, $E=0$.

We begin with the upper bound on the residual $E$ of a single pentagon or hexagon (the procedure treats each individually for simplicity). We want a single bound $|\sigma_u(E)|\leq H$ valid for every embedding 
\begin{equation}
    \sigma_u : K\hookrightarrow \mathbb C, ~~~~ \zeta\mapsto \zeta^u,~~~~u\in(\zee{}/n\zee{})^\times
\end{equation}
For a single $F$-symbol $x = \sum_k q_k \zeta^k \in K$ with $q_k\in\mathbb Q$, the desired upper bound is given by the $\ell_1$-norm of its coefficients
\begin{equation}
    |\sigma_u(x)|\leq \sum_k |q_k| = h(x)
\end{equation}
Then, by the triangle inequality, we may take $H$ to be the sum of the two sides of the defining equation of $E$, evaluated with $h(x)$ in place of each $F$-symbol $x$ and absolute values in place of the fixed $R$-symbols. The latter are zero or roots of unity, so their absolute values are zero or one in every embedding. This evaluation is simple, since it is performed in $\mathbb Q$ instead of $K$.

Next, we compute the ideal clearing the denominators of the residual $E$. For each $F$-symbol $x$, we compute its denominator ideal 
\begin{equation}
    \mathcal D_x = \{a\in \mathcal O_K:ax\in\mathcal O_K\}
\end{equation}
Then, substituting these into the defining equation of $E$ yields an integral ideal $\mathcal B$ such that $\mathcal B(E)\subset \mathcal O_K$. Its norm is $N(\mathcal B) = [\mathcal O_K:\mathcal B]$, and $\mathcal B(E)$ is a nonzero integral ideal if $E\neq 0$.

As in the earlier propagation, we evaluate the pentagons and hexagons modulo sufficiently many primes $p_i\equiv 1\pmod{n}$. We verify that $E$ vanishes modulo every prime ideal above each $p_i$, taking only $p_i$ coprime to $N(\mathcal B)$. Hence $\mathcal B(E) \subset (p_i)$ for each $i$, and so $\mathcal B(E)\subset (M)$ for $M = \prod_i p_i$. If $E\neq 0$, taking norms gives \begin{equation}M^{\varphi(n)} \leq N(\mathcal B(E)) = N(\mathcal B) |N_{K/\mathbb Q}(E)| \leq N(\mathcal B) H^{\varphi(n)}\end{equation} where in the last inequality we use that $N_{K/\mathbb Q}(E) = \prod_u \sigma_u(E)$ and $|\sigma_u(E)|\leq H$ for each $u$. Thus, once enough primes are used that $M > N(\mathcal B)^{1/\varphi(n)} H$, we arrive at a contradiction, and so $E = 0$. This procedure is applied to each pentagon and hexagon and provides a far cheaper route to certifying a solution than exact evaluation in $K$.

\section{Discussion}
\label{sec:disc}

We have described a general method for computing the $F$-symbols of unitary modular categories with a self-dual simple generator $X$. Large classes of modular categories have such a generator. For instance, such a generator exists for 1,669 of the 3,220 known pseudo-unitary modular data up to rank 22 in~\cite{Berean-Dutcher:2026fck}. Our motivating rank-11 example has the most $F$-symbols of any modular data at its rank~\cite{Berean-Dutcher:2026fck}, and more than twice as many as the Haagerup center $\mathcal Z(\mathcal H_3)$. We regard this example as encouraging evidence for scalability with both rank and multiplicity. The authors of~\cite{maurerFsymbolsRsymbolsDrinfeld2026} previously computed the $F$-symbols of $\mathcal Z(\mathcal H_3)$. With our algorithm, we can compute the $F$-symbols of either category in roughly five minutes.

Our algorithm still relies largely on brute-force numerical searches and will encounter computational limits in sufficiently large categories. It may fail to find existing solutions or miss additional solutions corresponding to modular isotopes, inequivalent modular categories with the same modular data~\cite{mignardModularCategoriesAre2021}. Moreover, it is not guaranteed that the constraints at every necessary rung suffice to fix the relevant $F$-symbols; rungs beyond $n=5$ may be needed if certain simple objects first occur in higher powers of the generator. The algorithm appears reliable at least through rank 12 and, when a solution is found and reconstructed, the computed $F$-symbols can always be verified exactly.

As we discuss in \cref{sec:assumptions}, we can likely relax the unitarity and cyclotomic-field assumptions. It may also be possible to replace $X$ by a generating pair of simples $X,Y$ and construct the ladder using braids on strands of both types. Whether one could carry out these generalizations at a tractable computational cost remains unclear.

The overall strategy does not depend on the choice of numerical solver, although a particularly sophisticated one was needed to implement this strategy in the rank-11 case. We intend to generalize our implementation before releasing a more polished version of the solver, and in the meantime will add further $F$-symbol solutions to \href{https://anyonatlas.org}{\texttt{AnyonAtlas.org}}.

\FloatBarrier

\vspace{5pt}
\noindent\textbf{Acknowledgements.}
We are grateful to \href{https://projectnumina.ai}{Project Numina} for computing support and collaboration on related topics. JBD is supported by a Four-Year Fellowship (4YF) from the University of British Columbia. CC and RT are supported by the Simons Collaboration on Global Categorical Symmetries and the US Department of Energy Grant 5-29073. The overall strategy was designed by the authors with the help of AI models. Its successful implementation benefited substantially from their use.


\begingroup
\emergencystretch=2em
\printbibliography
\endgroup

\end{document}